\documentclass[11pt]{article}

\usepackage{amsmath,amsfonts,amssymb,bm}

\usepackage[utf8]{inputenc}
\usepackage[T1]{fontenc}
\usepackage{hyperref}
\usepackage{graphicx,bm,amsmath,color}
\usepackage{bbold}
\usepackage{wasysym}
\usepackage{verbatim}
\usepackage{comment}
\usepackage[symbol]{footmisc}
\usepackage{epstopdf}
\usepackage{animate}
\usepackage{cite}
\usepackage{xmpmulti}
\usepackage{centernot}
\usepackage{multirow}
\usepackage{tikz}
\usepackage{graphicx}
\usepackage{float}
\usepackage{mathtools}
\usepackage{braket}

\makeatletter

\usepackage {amsmath}  
\newcommand{\be}{\begin{equation}}
\newcommand{\ee}{\end{equation}}
\newcommand{\beq}{\begin{eqnarray}}
\newcommand{\eeq}{\end{eqnarray}}
\newcommand{\ba}{\begin{align}}
\newcommand{\ea}{\end{align}}

\newcommand{\up}{\uparrow}
\newcommand{\down}{\downarrow}

\def\co{\Delta}  

\def\be{\begin{equation}}
\def\ee{\end{equation}}
\def\bea{\begin{eqnarray}}
\def\eea{\end{eqnarray}}

\def\conm#1#2{\left [ {#1},{#2} \right ]}

\def\ket#1{| #1 \rangle}

\begin{document}

\ \hfill \today

\begin{center}
\baselineskip 24 pt 
{\LARGE \bf  
Quantum group deformation of the Kittel--Shore model}
\end{center}

\bigskip
\bigskip
\begin{center}

{\sc A. Ballesteros$^{1}$, I. Gutierrez-Sagredo$^{2}$,  V. Mariscal$^{2}$ and J.J. Relancio$^{2,3}$*\footnotetext{*
Author to whom any correspondence should be addressed.}}

\medskip
{$^1$Departamento de F\'isica, Universidad de Burgos, 
09001 Burgos, Spain}

{$^2$Departamento de Matem\'aticas y Computaci\'on, Universidad de Burgos, 
09001 Burgos, Spain}

{$^3$Centro de Astropart\'{\i}culas y F\'{\i}sica de Altas Energ\'{\i}as (CAPA),
Universidad de Zaragoza, Zaragoza 50009, Spain}
\medskip
 
e-mail: {\href{mailto:angelb@ubu.es}{angelb@ubu.es}, \href{mailto:igsagredo@ubu.es}{igsagredo@ubu.es}, \href{mailto:vmariscal@ubu.es}{vmariscal@ubu.es}, \href{mailto:jjrelancio@ubu.es}{jjrelancio@ubu.es}}

\end{center}

\medskip

\begin{abstract}
The Kittel--Shore (KS) Hamiltonian describes $N$ spins with long-range interactions that are identically coupled; therefore, this (mean-field) model is also known as the Heisenberg XXX model on the complete graph. In this paper, the underlying $U(\mathfrak{su}(2))$ coalgebra symmetry of the KS model is demonstrated for arbitrary spins, and the quantum deformation of the KS Hamiltonian ($q$-KS model) is obtained using the corresponding $U_q(\mathfrak{su}(2))$ quantum group. By construction, the existence of such a symmetry guarantees that all integrability properties of the KS model are preserved under $q$-deformation. In particular, the $q$-KS model for spin-$1/2$ particles is analysed, the cases with $N=2$ and $3$ spins are studied in detail, and higher-spin $q$-KS models are sketched. {As a first excursion into the thermodynamic properties of the spin-$1/2$ $q$-KS model, the dependence of the Curie temperature on the deformation parameter is studied through numerical analysis}.
\end{abstract}

\medskip
\medskip

%\noindent
%PACS:   
%\bigskip


\tableofcontents

\section{Introduction}

Let us consider the KS Hamiltonian~\cite{kittel1965development} that describes a system of $N$ long-range interacting spins with identical coupling among all of them, namely
\begin{equation}
    H_{KS}=- I \sum_{i<j}^{N}\Vec{J}_{i}\cdot\Vec{J}_{j} \, ,
    \label{eq:KS}
\end{equation}
where $\Vec{J}_{i}=(J_{x}^{(i)},J_{y}^{(i)},J_{z}^{(i)})$ are the (arbitrary) spin operators for the $i$-th particle. The coupling constant distinguishes between ferromagnetic ($I>0$) and antiferromagnetic cases ($I<0$), respectively. When a constant external magnetic field $\vec{h}=(0,0,h)$ is considered, the Hamiltonian reads
\begin{equation}
    \tilde H_{KS}=- I \sum_{i<j}^{N}\Vec{J}_{i}\cdot\Vec{J}_{j}-\gamma \,h\, \sum_{i=1}^{N}J_{z}^{(i)},
    \label{eq:hamiltonian1}
\end{equation}
where $\gamma=g\mu_{B}$, in the usual notation. 

As Kittel and Shore emphasised in their original paper~\cite{kittel1965development}, it defines a `rigorously solvable' many-body system, whose exact solvability stems from the well-known representation theory of the $U(\mathfrak{su}(2))$ Lie algebra, particularly because of the well-known decomposition of the tensor product of $N$ irreducible representations of $U(\mathfrak{su}(2))$ into irreducible blocks. Consequently, the Bethe Ansatz technique is not required to solve this problem. Moreover, this system provides an exactly solvable framework to study the theory of phase transitions for long-range interacting systems; in particular, it was found to present a ferromagnetic phase transition that develops very slowly as $N$ increases~\cite{kittel1965development}. Although the spin-$1/2$ case was studied in~\cite{kittel1965development} and later in~\cite{al1998exact,van1993heisenberg,czachor2002verification}, the underlying integrability properties of the model are preserved for arbitrary spins. The study of the KS model was extended to the antiferromagnetic case in~\cite{al1998exact,czachor2008energy}.

{It is worth stressing that the KS Hamiltonian is just the Heisenberg XXX model on the complete graph (see~\cite{bjornberg2020quantum} and references therein), {\em i.e.}~the long-range generalisation for arbitrary spin of the Heisenberg XXX model with a fully symmetric constant coupling among all spins, and provides a canonical example of a mean-field Hamiltonian~\cite{fannes1980equilibrium} (provided that $I$ is transformed into $I/N$). Quantum spin models on the complete graph present a number of outstanding analytical features, that have been deeply analysed for the spin-1/2 and spin-1 cases. For instance, an integral formula for the partition function in the thermodynamic limit (for the ferromagnetic case), explicit expressions for the magnetization, free energy and critical exponents can be found (see~\cite{van1993heisenberg,bjornberg2020quantum,ryan2023class,toth1990phase,penrose1991bose,bjornberg2016free,alon2021mean,bjornberg2023heisenberg} and references therein). In the spin-1 case, the properties of a KS model as a bilinear biquadratic Hamiltonian have also been analysed~\cite{papanicolaou1986ground,jakab2018bilinear}. In general, the underlying rotational and permutational symmetries of the model play a relevant role in most of the abovementioned results.
Therefore, the $H_{KS}$ Hamiltonian} can be viewed as a model of $N$ spins located on the vertices of a $(N-1)$-dimensional simplex and, in this context, the model can be considered as infinite dimensional in the thermodynamic limit $N\to \infty$. Nevertheless, the cases with $N=2,3,4$ (dimer, equilateral triangle, and regular tetrahedron) arrays of spins have been realised in the context of ultra-small magnetic clusters (see~\cite{ciftja1999equation} and references therein) or as Hamiltonians for quantum computation with quantum dots~\cite{loss1998quantum,woodworth2006few}.

The KS model for classical spins has also been studied systematically from different viewpoints as a long-range integrable generalisation of the Heisenberg model, the dynamical behaviour of which, including the KS case, was analysed in \cite{liu1991dynamics}. The $N$ independent integrals in the involution of the KS model, as well as its action-angle variables, were presented in~\cite{magyari1987integrable}, and the autocorrelation function of the so-called `equivalent-neighbour Heisenberg model' (classical KS model) was calculated analytically in~\cite{muller1988high}. As another example of its applications, in~\cite{botet1982size,botet1983large}, the KS model was used to demonstrate the extension of Fisher's and Barber's finite-size scaling hypothesis~\cite{cardy1988current} to the case of `infinitely coordinated systems' (long-range interactions). Moreover, the correlation functions for the anisotropic KS model (the so-called `equivalent-neighbour XYZ model') have also been considered ~\cite{dekeyser1979time,lee1984time}.

On the other hand, it is also well known that quantum groups~\cite{vladimir1986drinfeld,jimbo1985q} provide symmetries of certain integrable spin chains endowed with specific boundary conditions. This is the case for the XXZ Heisenberg model for $N$ spin-$1/2$ particles, for which $U_q(\mathfrak{su}(2))$ turns out to be the symmetry algebra, provided that the quantum deformation parameter $q$ and the anisotropy parameter $\delta$ of the model are related through $\delta=(q+q^{-1})/2$, and the extra boundary term is just $(q-q^{-1})(\sigma_z^{1} - \sigma_z^{N})/2$ (see~\cite{sklyanin1988boundary,pasquier1990common,gomez1996quantum} and references therein). In this sense, the spin-$1/2$ Heisenberg XXZ Hamiltonian can be understood as a $U_q(\mathfrak{su}(2))$ deformation of the isotropic XXX model, which is recovered in the limit $q\to 1$ because under this condition, both the anisotropy parameter $\delta$ and boundary term vanish. {Note that the periodic Heisenberg XXZ model can also be considered as a $q$-deformation of the Heisenberg XXX Hamiltonian in the more standard sense of quantum integrable models, since the whole family of the quantum integrals of the motion can be shown to be $U_q(\mathfrak{su}(2))$ invariant~\cite{kulish1991general}. Moreover, other spin models endowed with $U_q(\mathfrak{su}(2))$ symmetry are known. These include  $q$-deformations of long-range interacting models such as the `braid translate'~\cite{martin1993algebraic,martin1994blob} $q$-deformed periodic Heisenberg XXZ model and the $q$-deformed Haldane-Shastry model (see~\cite{lamers2022spin} and references therein), as well as systems with quantum affine algebra symmetry~\cite{hakobyan1996spin},  elliptic generalizations of the $q$-deformed Haldane-Shastry model~\cite{matushko2022elliptic} and integrable deformations of Inozemtsev spin chains~\cite{klabbers2024deformed}. It is worth stressing that these models present nonlocal interactions induced by the $q$-deformation, a feature that will also arise as a distinctive feature in the model here introduced.}

%\red{Several papers in the literature have been used spin models with $U_q(\mathfrak{su}(2))$ symmetry. This is the case of the the  $q$-deformed periodic Heisenberg XXZ model \cite{martin1993algebraic,martin1994blob,lamers2022spin}     and $q$-deformed Haldane--Shastry model \cite{uglov1995trigonometric,hakobyan1996spin,lamers2018resurrecting,lamers2022spin}. Moreover, this symmetry has been appeared in recent years in some $q$-deformed integrable long range models, such as the $q$-deformations of the anisotropic Haldane--Shastry~\cite{matushko2022elliptic} and Inozemtsev spin chains  \cite{klabbers2024deformed}, which all have similarly looking nonlocal interactions.}

The aim of this paper is to explicitly present and solve the $U_q(\mathfrak{su}(2))$ analogue of the KS model~\eqref{eq:hamiltonian1} for $q$ that is not a root of unity. {The significance of such $q$-KS model is two-fold: from the mathematical viewpoint, it is a new instance of a quantum group invariant spin system with long-range interactions which is completely integrable for any spin. From the physical perspective, the $q$-KS model provides a smooth one-parametric deformation of the KS model containing site-dependent interactions  which could be of interest wherever the KS model turns out to be relevant as, for instance, dealing with systems composed of a small number of spins.} 
The definition of this new quantum-group invariant model will be possible after unveiling the $U(\mathfrak{su}(2))$ coalgebra symmetry~\cite{ballesteros1998systematic} of the undeformed KS Hamiltonian, which will be presented in the next Section. Thus, the KS model will also be revisited from a coalgebraic perspective: its eigenvalues and eigenvectors will be derived, thus allowing the exact derivation of its partition function. It is also worth noting that the Zeeman effect induced by the magnetic field is not affected by the $q$-deformation. 

The $q$-deformation of the KS model is introduced in Section 3 by imposing its $U(\mathfrak{su}(2))$ coalgebra symmetry. In particular, the $q$-KS Hamiltonian for $N$ spin-$1/2$ particles is thoroughly studied in Section 4. As essential results, we found that owing to the underlying coalgebra symmetry, the $q$-KS Hamiltonian preserves the complete integrability of the KS model and presents the same degeneracies of its spectrum, which nevertheless becomes nonlinearly deformed with respect to that of the KS model. These results are based on the fact that, where $q$ is not a root of unity, the representation theory of $U_q(\mathfrak{su}(2))$ is structurally identical to that of $U(\mathfrak{su}(2))$~\cite{biedenharn1995quantum}. Moreover, to realise the $U_q(\mathfrak{su}(2))$ invariance of the Hamiltonian, the $q$-KS model is forced to break the characteristic KS symmetry between the interactions among all the $N$ spins, which now become quite involved and site dependent, a fact that is induced by the non-local nature of the $U_q(\mathfrak{su}(2))$ coproduct maps. Nevertheless, the underlying quantum group symmetry allows for the explicit knowledge of the full spectrum and eigenvectors, from which the partition function can be derived. {Moreover, from the explicit expressions of the model for the spin-$1/2$ case, it will become clear that the $q$-KS model is by no means equivalent to the XXZ Heisenberg chain on the complete graph. Therefore, we have obtained a new exactly solvable $N$-dimensional quantum model on the complete graph for arbitrary spin and with site-dependent long-range interactions.}

In particular, the density of states of the $q$-KS model {for $q\in  {\rm I\!R}$} will be compared with that of the KS model, and the smooth effect of the deformation can be appreciated. A deeper analysis of the $q$-KS Hamiltonian {for $N=2$ and $3$} spin-$1/2$ particles is provided, including the explicit form of the eigenvectors, which can be obtained through the iterated application of the $q$-Clebsch--Gordan coefficients~\cite{biedenharn1995quantum,alvarez2024analog}. In the ferromagnetic case, the ground-state eigenvectors are the so-called $q$-Dicke states~\cite{li2015entanglement, raveh2024q, qpermutations}, which are the most excited states in the antiferromagnetic model. The ground states for the antiferromagnetic $q$-KS model are also presented, which conversely provide the highest excited states for the ferromagnetic case. In Section 5, the $q$-deformed KS model for $N$ spin-$1$ particles is explicitly presented, and the $q$-KS models with arbitrary spin are sketched using the so-called functional realisation approach to the $U_q(\mathfrak{su}(2))$ representations~\cite{curtright1990deforming,fairlie1990quantum,ballesteros1992characterization}. {The study of the Curie temperature  for the spin-$1/2$ $q$-KS model is faced in Section 6, where new numerical strategies have to be developed because of the lack of complete analytical solutions for the $q$-deformed analogues of the equations defining $T_C$ in the KS model. The results here presented show the existence of the corresponding phase transition for the magnetization and provide the basis for the complete study of the thermodynamics of the $q$-KS model, {which should be affected by the significant changes that the $q$-deformed partition function undergoes for different values of the deformation parameter and will reflect the physical consequences of the strongly nonlocal interactions introduced in the model by the $q$-deformation.} 
Finally, a Section that includes some remarks and open problems closes this paper.

\section{The Kittel-Shore model}

In this Section, we review the well-known KS model, unveiling its $U(\mathfrak{su}(2))$ coalgebra symmetry. This fact will be shown to have profound implications for the complete solvability of the model and for the possibility of defining an integrable $q$-deformation of the KS Hamiltonian, which will be presented in the next Section. 

{In what follows we will consider the Hamiltonians~\eqref{eq:KS} and \eqref{eq:hamiltonian1} as `abstract' objects defined in terms of $N$ copies of the generators of the $U(\mathfrak{su}(2))$ Lie algebra, which} satisfy the following commutation relations
\begin{equation}
[J_x,J_y]= i J_z, \qquad   [J_y,J_z]= i J_x, \qquad   [J_z,J_x]= i J_y.
\end{equation}
The Casimir operator for this Lie algebra is given by
\begin{equation}
C= J_x^2 +J_y^2 + J_z^2 \, .
\end{equation}
If we perform the change of basis 
\begin{equation}
    J_\pm = J_x \pm i J_y ,
\qquad
\qquad
    J_x = \frac{1}{2} (J_+ + J_-), \qquad J_y = - \frac{i}{2} (J_+ - J_-)\, ,
\end{equation}
the commutation rules become
\begin{equation}
[J_z,J_\pm]= \pm J_\pm,\qquad [J_+J_-]= 2\,J_z \, ,
\label{lpm}
\end{equation}
and the Casimir operator now reads
\begin{equation}
C= J_- J_+ +J_z^2 + J_z=J_+ J_- + J_z^2- J_z\, .
\end{equation}

As it is well-known, the spin-$j$ irreducible representation $D^j$ of the $U(\mathfrak{su}(2))$ algebra is defined on a $(2j+1)$-dimensional space by the action on the $|j, m\rangle$ basis vectors given by
\begin{equation}
J_\pm |j, m\rangle = \sqrt{(j\mp m) (j\pm m+1)}|j ,m\pm 1\rangle,\qquad  J_z |j, m\rangle=m |j, m\rangle\, ,
\label{irepsu2}
\end{equation}
where $j=0,1/2, 1, 3/2, \dots$ and $m=-j,\dots,+j$. The common eigenvalue of the Casimir operator on all the $D^j$ basis vectors is given by $C |j, m\rangle=j(j+1) |j, m\rangle$ and provides the expected value of the squared modulus of the spin.

The tensor product representation $D^{j_1}\otimes D^{j_2}$ turns out to be fully reducible, and it can be decomposed into irreducible components in the form
\begin{equation}
D^{j_1+j_2}\oplus D^{j_1+j_2 -1} \oplus \dots\oplus D^{|j_1-j_2|} \, .
\label{blockd}
\end{equation}
The Clebsch--Gordan coefficients provide the transformation matrix between the $((2j_1+1)(2j_2+1))$-dimensional  basis $|j_1, m_1\rangle\otimes |j_2, m_2\rangle$ corresponding to the tensor product representation $D^{j_1}\otimes D^{j_2}$ and the block-diagonal one~\eqref{blockd}, whose basis will be given by the vectors
\begin{equation}
\{|j_1+j_2, m_{j_1+j_2}\rangle, |j_1+j_2 -1, m_{j_1+j_2-1}\rangle , \dots , 
||j_1-j_2|, m_{|j_1-j_2|}\rangle \}\, , \qquad m_\alpha=-j_\alpha,\dots,j_\alpha \, ,
\label{blockdbasis}
\end{equation}
where subscript $\alpha$ labels each irreducible subspace.

The iteration of this procedure allows the transformation of the tensor product representation of $N$ arbitrary irreducible representations of the $U(\mathfrak{su}(2))$ algebra 
\begin{equation}
D^{j_1}\otimes D^{j_2}\otimes \dots\otimes D^{j_N} \, ,
\label{directsum}
\end{equation}
into its block-diagonal form in terms of $k$ irreducible representations
\begin{equation}
D^{j_{\alpha_1}}\oplus D^{j_{\alpha_2}} \oplus \dots\oplus D^{j_{\alpha_k}} \, ,
\label{reduced}
\end{equation}
where $j_{\alpha_1}=j_1+j_2+\dots + j_N$ is the spin of the representation with the highest dimension and $(j_{\alpha_2}, j_{\alpha_3}, \dots, j_{\alpha_k})$ are the spins of each of the lower-dimensional irreducible representations that arise in the decomposition. In general, some of these spins $j_{\alpha_i}$ can be the same, giving rise to the so-called multiplicities for a given irreducible representation with spin $J$ (see~\cite{curtright2017spin} and references therein for the derivation of explicit formulas giving the multiplicities when the tensor product~\eqref{directsum} of $N$ representations with identical spin $j$ are considered). 
The associated block-diagonal basis for the representation~\eqref{reduced} will be given by the vectors
\begin{equation}
\{|j_{\alpha_1}, m_{\alpha_1}\rangle, |j_{\alpha_2}, m_{\alpha_2}\rangle, 
\dots, |j_{\alpha_k}, m_{\alpha_k}\rangle\}\, , \qquad m_{\alpha_i}=-j_{\alpha_i},\dots,j_{\alpha_i} \, ,
\end{equation}
that can be expressed in terms of the initial basis vectors corresponding to~\eqref{directsum}
\begin{equation}
|j_1, m_1\rangle\otimes |j_2, m_2\rangle\otimes\dots\otimes |j_N, m_N\rangle \, , \qquad m_{i}=-j_{i},\dots,j_{i} \, , \qquad i=1,\dots, N\, ,
\end{equation}
using the corresponding iteration of the Clebsch--Gordan transformations. Explicit examples of this will be provided in the following Sections.

\subsection{Coalgebra symmetry of the KS model}

A Hopf algebra (see, for instance~\cite{ChariPressley}) is a (unital, associative) algebra $A$ endowed with two homomorphisms called coproduct $(\Delta : A\longrightarrow A\otimes A )$ and counit $(\epsilon : A\longrightarrow \mathbb C)$, as well as an antihomomorphism (the antipode $S : A\longrightarrow A$) such that,
$\forall a\ \! \in A$:
\bea
&(id\otimes\co)\co (a)=(\co\otimes id)\co (a),  
\label{co}\\
&(id\otimes\epsilon)\co (a)=(\epsilon\otimes id)\co (a)= a, 
\label{coun}\\
&m((id\otimes S)\co (a))=m((S \otimes id)\co (a))=
\epsilon (a) 1, 
\label{ant}
\eea
where $m$ is the usual multiplication mapping $m(a\otimes b)=a\cdot b$ and $id:A\rightarrow A$ is the identity map. The condition~\eqref{co} is called the coassociativity property of $\co$. When $A$ is considered as a module endowed with comultiplication and counit maps $(\co,\epsilon)$ fulfilling~\eqref{co} and~\eqref{coun}, then $A$ is said to be endowed with a coalgebra structure. 
 
This means that the coproduct map $\co$ underlies the construction of the tensor product representation of two copies of the Hopf algebra $A$. Moreover, the coassociativity (\ref{co}) of the coproduct map $\co$ provides its generalisation to any number of copies of $A$.  If we denote $\co^{(2)}\equiv\co$, then  the homomorphism $\co^{(3)}:A\rightarrow
A\otimes A\otimes A$ can be defined using any of the two equivalent expressions:
\be \co^{(3)}:=(id\otimes\co^{(2)})\circ\co^{(2)}=(\co^{(2)}\otimes id)\circ\co^{(2)}\, ,
\label{fj}
\ee
and this construction
can be generalized to an arbitrary number $N$ of tensor products of $A$ either by the
recurrence relation {
\be
\co^{(N)}:=(id^{\otimes(N-2)} \otimes
\co^{(2)})\circ\co^{(N-1)},
\label{fl}
\ee}
or by the equivalent one 
{\be
\co^{(N)}:=(\co^{(2)}\otimes id^{\otimes(N-2)})\circ\co^{(N-1)}.
\label{fla}
\ee}

In particular, the universal enveloping algebra $\mathcal U (\mathfrak{g})$ of any Lie algebra $\mathfrak{g}$ can be endowed with a Hopf algebra structure by defining
 \be
 \co_0(X)=\mathbf{I}\otimes X + X\otimes \mathbf{I}, \quad\forall X\in\mathfrak{g}\qquad
\co_0(\mathbf{I})=\mathbf{I}\otimes \mathbf{I},\label{coprim} 
\ee
where $\mathbf{I}$ denotes the identity element in $\mathcal U (\mathfrak{g})$. The coproduct map can be extended
to any monomial in $\mathcal U (\mathfrak{g})$ by means of the homomorphism condition for the coproduct
$\co_0(X\cdot Y)=\co_0(X)\cdot \co_0(Y)$, which implies the compatibility of the coproduct
$\Delta$ with the Lie bracket
\be
\conm{\co_0(X_i)}{\co_0(X_j)}_{\mathfrak{g}\otimes \mathfrak{g}}=\co_0(\conm{X_i}{X_j}_\mathfrak{g}),\qquad \forall
X_i,X_j\in \mathfrak{g}. \label{hom}
\ee
From~\eqref{fl} is straightforward to prove that the $N$-th coproduct of the Lie algebra generators is given by
{
\bea
&&\!\!\!\!\!\!\!\!\co_0^{(N)}(X_i)=X_i\otimes \mathbf{I}^{\otimes(N-1)}  + \mathbf{I}\otimes X_i\otimes\mathbf{I}^{\otimes(N-2)} +
\dots + \mathbf{I}^{\otimes(N-1)}\otimes X_i.
\label{coN}
\eea}
Let us recall that a Hamiltonian $\mathcal{H}$ is said to be endowed with the so-called ``coalgebra symmetry''~\cite{ballesteros1998systematic,BCR} if it can be written as a function of the coproduct map {$\Delta^{(N)}$} of the generators of a given coalgebra $A$. 

Let us now consider $N$ copies of the  $A\equiv \mathfrak{su}(2)$ Lie algebra~\eqref{lpm} with generators $\{J_{\pm}^{(i)}, J_{z}^{(i)}\}$, where the $N$-th coproduct map~\eqref{coN} of the $U(\mathfrak{su}(2))$ generators provide the collective spin operators 
 \begin{equation}
\Delta_0^{(N)}(J_{\pm})=  \sum_{i=1}^N J_{\pm}^{(i)}, \qquad \Delta_0^{(N)}(J_{z})=  \sum_{i=1}^N  J_z^{(i)},
\label{eq:coproduct_nsd}
\end{equation}
and superindex $i$ represents the spin operator of the $i$-th particle. By making use of these expressions, it can be straightforwardly proven that the KS Hamiltonian~\eqref{eq:hamiltonian1} can be rewritten as
\begin{equation}
    \tilde H_{KS}=- I \sum_{i<j}^{N}\Vec{J}_{i}\cdot\Vec{J}_{j}-\gamma \,h\, \sum_{i=1}^{N}J_{z}^{(i)}= -\frac{I}{2}\left(\Delta_0^{(N)}(C)-\sum_{i=1}^N C^{(i)}\right)-\gamma h\Delta_0^{(N)}(J_{z}),
    \label{KSco}
\end{equation}
where
$
C^{(i)}=J_-^{(i)}\,J_+^{(i)}+ (J_z^{(i)})^2 + J_z^{(i)}= (J_{x}^{(i)})^{2}+(J_{y}^{(i)})^{2}+(J_{z}^{(i)})^{2}
$
is the Casimir operator for the $i$-th $U(\mathfrak{su}(2))$ algebra. Equation~\eqref{KSco} is proven by making use of the expressions~\eqref{eq:coproduct_nsd} and by expanding the coproduct $\Delta_0^{(N)}(C)$ as $\Delta_0^{(N)}(J_- J_+ +(J_z)^2+J_z)$, namely
\begin{equation}
    \tilde H_{KS}=-\frac{I}{2} \left(\Delta_0^{(N)} (J_{-})\Delta_0^{(N)} (J_{+})+(\Delta_0^{(N)} (J_{z}))^2 + \Delta_0^{(N)} (J_{z})-\sum_{i=1}^N C^{(i)}\right)-\gamma h \Delta_0^{(N)}(J_{z}) \, .
\label{KScoalgebra}
\end{equation}
{Afterwards, when a matrix representation of~\eqref{KScoalgebra} is to be obtained, different irreps $j_i$ could be considered for each lattice site.}

The expression~\eqref{KScoalgebra} implies that the KS model is endowed with coalgebra symmetry~\cite{ballesteros1998systematic} (see also~\cite{BCR,ballesteros2003classical,ballesteros2009super}, and references therein), as the Hamiltonian can be written as a function of the coproducts of the generators of a given Hopf algebra, in this case $\mathcal U (\mathfrak{su}(2))$ {(note that $\sum_{i=1}^N C^{(i)}$ obviously  belongs to the centre of $\mathfrak{su}(2)^{\otimes N}$ and will give rise to a constant that depends on the representation considered in each case)}. As it was demonstrated in~\cite{ballesteros1998systematic}, the complete integrability of any coalgebra-symmetric system is a consequence of the fact that for any coalgebra-symmetric Hamiltonian $H^{N}$ the coproducts $\co^{(m)}(C)$ with $m=2,\dots,N$ of the Casimir operators provide $(N-1)$ constants of the motion in involution, namely
\be
\conm{H^{(N)}}{\co^{(m)}(C)}=0,
\qquad
\conm{\co^{(m)}(C)}{\co^{(k)}(C)}=0,
\qquad \forall\, m,k=2,\dots, N \, .
\label{cointegrability}
\ee
{Here the previous commutators are defined on $U (\mathfrak{su}(2))^{\otimes(N)}$, and we also understand that $\co^{(m)}(C)\equiv \co^{(m)}(C)\otimes\mathbf{I}^{\otimes(N-m)}$ for any $m=2,\dots,N$. }
 
 {The functional independence among the integrals given by $\co^{(m)}(C)$ is guaranteed by the fact that for each $m=2,\dots,N$ this object is defined on a different number of copies of  $U(\mathfrak{su}(2))$. In addition, it is worth stressing that the non-triviality of the coproduct $\co^{(m)}(C)$ of a given element $C\in U(A)$ is ensured whenever $C$ is a nonlinear function of the generators of $A$. For other Lie algebras where $C$ is a linear function, a coproduct of the type~\eqref{coN} would give rise to a set of integrals that reduce to numerical constants under any irreducible representation (see~\cite{ballesteros1998systematic,ballesteros2009super} for more detailed discussions).}

Therefore, for any Hopf algebra $A$, the previous result enables the construction of a large class of integrable systems, which can be in principle diagonalised on a basis constructed from common eigenvectors of the $\co^{(m)}(C)$ operators (see, for instance, the case of different $\mathfrak{sl}(2)$ classical and quantum Calogero-Gaudin models and their $q$-deformations~\cite{ballesteros1998systematic,musso1999exact,musso2000spin,musso2005gaudin}). In the case of the KS model presented here, the basis of $\co^{(m)}(C)$ eigenvectors is not needed because the block-diagonal one~\eqref{blockdbasis} provides eigenvectors and eigenvalues in a straightforward manner, as we shall see below.

More importantly, a second essential consequence of the coalgebra symmetry of a given Hamiltonian $H$ defined on a Hopf algebra $A$ is that any quantum deformation $A_q$ of the underlying Hopf algebra structure automatically induces an integrable deformation $H_q$ of the initial Hamiltonian~\cite{ballesteros1998systematic}. Such a deformed Hamiltonian is defined by substituting the coproduct of the generators with its quantum deformation, both in the definition of the Hamiltonian and in the $m$-th order coproducts of the $q$-deformed Casimir $C_q$. This is just the procedure that will be used in the next Section to construct a quantum deformation of the KS model by using the quantum deformation of  $\mathcal U(\mathfrak{su}(2))$.

\subsection{Eigenvectors and eigenvalues}

Let us firstly consider the KS Hamiltonian~\eqref{eq:KS} without external magnetic field ($h=0$) in the coalgebra-symmetric form~\eqref{KSco}, namely
\begin{equation}
    H_{KS}=-\frac{I}{2}\left(\Delta_0^{(N)}(C)-\sum_{i=1}^N C^{(i)}\right)\, ,
\label{KScoalgebrah0}
\end{equation}
where we recall that $C$ denotes the $U(\mathfrak{su}(2))$ Casimir operator~\eqref{lpm}. Since the coproduct map $\Delta_0^{(N)}$ is an algebra homomorphism, we have that
\be
\conm{\Delta_0^{(N)}(C)}{\Delta_0^{(N)}(X)}=\Delta_0^{(N)}(\conm{C}{X})=0\, ,
\qquad \forall\,  X \in U(\mathfrak{su}(2)).
\label{schur}
\ee
The previous expression implies, in particular, that the collective spin operators~\eqref{eq:coproduct_nsd} commute with the KS Hamiltonian, which is essentially the $N$-th coproduct of the Casimir operator for $U(\mathfrak{su}(2))$. Moreover, the total spin component in the $z$ direction $\Delta_0^{(N)}(J_{z})=  \sum_{i=1}^N  J_z^{(i)}$ can be considered a good quantum number for the eigenstates of the model.  

Therefore, we consider the block-diagonal basis~\eqref{blockdbasis} for the tensor product of $N$ irreducible representations of $U(\mathfrak{su}(2))$. 
Thus, the $\Delta_0^{(N)}(C)\equiv C^{(N)}$ operator will be represented as a block-diagonal matrix. 
Within each of the irreducible blocks with basis vectors
\be
|j_{\alpha_i}, m_{\alpha_i}\rangle,  \qquad m_{\alpha_i}=-j_{\alpha_i},\dots,j_{\alpha_i},
\label{basisbd}
\ee
relations~\eqref{schur} imply that $C^{(N)}$ commutes with the representation of all $U(\mathfrak{su}(2))$ generators and, as a consequence of Schur's lemma, within each irreducible block the operator $C^{(N)}$ has to be proportional to the identity.  In fact, the proportionality factor will be simply the $U(\mathfrak{su}(2))$ Casimir eigenvalue within such a block, that is, $j_{\alpha_i}(j_{\alpha_i}+1)$. Therefore, we have proven that
\be
\Delta_0^{(N)}(C)\,|j_{\alpha_i}, m_{\alpha_i}\rangle = j_{\alpha_i}(j_{\alpha_i}+1)\, |j_{\alpha_i}, m_{\alpha_i}\rangle \,,
\qquad
\forall \alpha_i=1,\dots,k
\, ,
\qquad
m_{\alpha_i}=-j_{\alpha_i},\dots,j_{\alpha_i} \, .
\ee
Finally, since the term $\sum_{i=1}^N C^{(i)}$ acting on the initial tensor product basis 
\be
|j_1, m_1\rangle\otimes |j_2, m_2\rangle \otimes \dots \otimes  |j_N, m_N\rangle \, ,
\label{tpbasis}
\ee
is just a constant, namely, $\sum_{i=1}^N j_i(j_i+1)$,  this term will not be affected by the consecutive Clebsch--Gordan transformations leading to the block-diagonal basis. Thus, we have proven that the eigenvectors of the KS Hamiltonian~\eqref{KScoalgebrah0} are just the basis vectors~\eqref{basisbd} of the block-diagonal basis, and their eigenvalues are
\be
H_{KS} |j_{\alpha_i}, m_{\alpha_i}\rangle=-\frac{I}{2}\left(\Delta_0^{(N)}(C)-\sum_{i=1}^N C^{(i)}\right)\,|j_{\alpha_i}, m_{\alpha_i}\rangle =
-\frac{I}{2}\left( j_{\alpha_i}(j_{\alpha_i}+1)-\sum_{i=1}^N j_i(j_i+1)\right)\,|j_{\alpha_i}, m_{\alpha_i}\rangle \, .
\ee
Finally, when the magnetic field $h$ is considered we have that, within each irreducible block and after the Clebsch--Gordan transformation is performed onto the $\Delta_0^{(N)}(J_{z})$ operator
\be
-\gamma h \Delta_0^{(N)}(J_{z})  |j_{\alpha_i}, m_{\alpha_i}\rangle =-\gamma h \,J_{z}^{(\alpha_i)}  |j_{\alpha_i}, m_{\alpha_i}\rangle =-\gamma\,h\, m_{\alpha_i} |j_{\alpha_i}, m_{\alpha_i}\rangle \, .
\ee
Summarizing, the eigenvalues of the KS model with magnetic field for arbitrary spins are given by
\be
\tilde E(j_{\alpha_i},m_{\alpha_i},K)=-\frac{I}{2}\left( j_{\alpha_i}(j_{\alpha_i}+1)-K\right) -\gamma\,h\, m_{\alpha_i} \, ,
\label{eigenvaluesKS}
\ee
where $\alpha_i$ (with $i=1,\dots,k$) labels the $k$ irreducible blocks of the completely reduced form of the $D^{j_1}\otimes D^{j_2}\otimes \dots\otimes D^{j_N}$ representation, where $m_{\alpha_i}=-j_{\alpha_i},\dots,+ j_{\alpha_i}$ and $K=\sum_{i=1}^N j_i(j_i+1)$. 

We stress that this result can be considered a consequence of~\eqref{cointegrability} because $H_{KS}$ with $h=0$ is proportional to $\co^{(N)}(C)$ and its eigenvalues $E(j_{\alpha_i},K)$ depend only on the modulus of the total spin for each lattice site. Moreover,  $\co^{(N)}(C)$ commutes with the collective operators~\eqref{eq:coproduct_nsd} that represent the three components of the total spin, all of which are conserved when $h=0$. When the magnetic field is turned on, the isotropy of the model is broken, and only $\Delta_0^{(N)}(J_{z})$ remains as a constant of the motion; therefore, its eigenvalue $m_{\alpha_i}$ provides the second quantum number for the eigenstates $|j_{\alpha_i}, m_{\alpha_i}\rangle$ of the model.

\subsection{The spin-$1/2$ model}

When all the $N$ particles carry the same $j=1/2$ irreducible representation, the spin operators can be written as
\begin{equation}
    J_x = \frac{1}{2} \sigma_x, \qquad J_y = \frac{1}{2} \sigma_y, \qquad J_z = \frac{1}{2} \sigma_z,
    \label{eq:angular}
\end{equation}
where $\sigma_i$ are the Pauli matrices 
\begin{equation}
\sigma_x = 
\begin{pmatrix}
0 & 1 \\
1 & 0 \\
\end{pmatrix}, \qquad
\sigma_y = 
\begin{pmatrix}
0 & -i \\
i & 0 \\
\end{pmatrix}, \qquad
\sigma_z = 
\begin{pmatrix}
1 & 0 \\
0 & -1 \\
\end{pmatrix}.
\label{eq:pauli}
\end{equation}
Therefore, the matrix form of the $\{J_+,J_-,J_z\}$ operators reads
\begin{equation}
J_+ =  J_x + i \, J_y=
\begin{pmatrix}
0 & 1 \\
0 & 0 \\
\end{pmatrix}
= \,\sigma_+ \, , \qquad
J_- =   J_x - i \, J_y=
\begin{pmatrix}
0 & 0 \\
1 & 0 \\
\end{pmatrix}
= \,\sigma_- \, , \qquad
J_z = \frac{1}{2} 
\begin{pmatrix}
1 & 0 \\
0 & -1 \\
\end{pmatrix} \, ,
\label{paulipm}
\end{equation}
and in this case the eigenvalues~\eqref{eigenvaluesKS} read
\be
\tilde E(j_{\alpha_i},m_{\alpha_i},3N/4)=-\frac{I}{2}\left( j_{\alpha_i}(j_{\alpha_i}+1)- 3\, N/4\right) -\gamma\,h\, m_{\alpha_i} \, ,
\qquad
i=1,\dots,k \, ,
\label{eigenvaluesKS12}
\ee
where the irreducible blocks arising from the decomposition of the {$(D^{1/2})\,^{\otimes(N)}$} representation are given by spins $j_{\alpha_i}$ that can take the values
\be
J=N/2, N/2-1, \dots, \delta \, ,
\ee 
with $\delta=0$ for $N$ even and $\delta=1/2$ for $N$ odd. In general, the multiplicities $ d_{JNj}$ of the irreducible representation with spin $J$ within the tensor product of $N$ copies of spin $j$ representations are explicitly given by~\cite{VanVleck} 
\begin{equation}
    d_{JNj}=\Omega(J,N,j)-\Omega(J+1,N,j), \quad \text{with} \quad \Omega(J,N,j)\equiv\text{coefficient of }x^{J}  \text{ in } 
    \left(x^j+x^{j-1}+\cdots +x^{-j}\right)^N.
\end{equation}
As mentioned in~\cite{curtright2017spin}, this expression can be written in a simplified way
\begin{equation}
    \Omega(J,N,j)=\sum_{k=0}^{\min\left(N,\lfloor\frac{Nj+J}{2j+1}\rfloor\right)}(-1)^{k}\begin{pmatrix}
        N\\
        k
    \end{pmatrix}\begin{pmatrix}
        Nj+J-(2j+1)k+N-1\\
        Nj+J-(2j+1)k
    \end{pmatrix} \, ,
\end{equation}
where  the floor function $\lfloor x\rfloor$ is the function that takes as input a real number $x$ and gives as output the greatest integer less than or equal to $x$.

For the spin-$1/2$ case, these coefficients are explicitly given by~\cite{VanVleck}
\begin{equation}
    d_{JN\frac12}= \Omega(J,N,1/2)-\Omega(J+1,N,1/2)=\binom{N}{N/2-J}-\binom{N}{N/2-(J+1)}\, ,
\label{degeneracies}
\end{equation}
where $0\leq J\leq N/2$ and $J=0$ is allowed only in the case with even $N$.

In the original KS paper~\cite{kittel1965development} without an external magnetic field, the energy spectrum~\eqref{eigenvaluesKS12} is modified in the form
\be
 E'(j_{\alpha_i},3\, N/4)=-\frac{I}{2}\left( j_{\alpha_i}(j_{\alpha_i}+1)- 3\, N/4\right) - E_0(N) \, ,
 \quad {\mbox{with}} \quad {E_0(N)=\left\{\begin{matrix}
     \frac{3IN}{8} \hspace{1.8cm} I<0 \\
     -\frac{IN}{8}(N-1) \hspace{0.3cm} I>0
 \end{matrix}\right.}
\label{corrE}
\ee
where the constant term $E_0(N)$ denotes the ground-state energy. In this way, the corrected ground-state energy $E'$ is always zero, independent of the number $N$ of the spins. In the ferromagnetic case, $E_0(N)$ corresponds to the eigenvalue of the block with the highest spin $J=N/2$, and in the antiferromagnetic case, to the block with the lowest spin $J=\delta$. When the magnetic field is introduced, we will have
\be
E'(j_{\alpha_i},m_{\alpha_i},3 N/4))=-\frac{I}{2}\left( j_{\alpha_i}(j_{\alpha_i}+1)- 3\, N/4\right) - E_0(N) -\gamma\,h\, m_{\alpha_i}  \, .
\label{corrEmag}
\ee

We will also adopt this approach because in this way we can compare the effect of the $q$-deformation  in the model more easily. Also, in~\cite{kittel1965development} the spectrum for the ferromagnetic case was written in terms of a quantum number {$p=0,1,2,\dots, p_m$ (where $p_m=N/2$ if $N$ is even and $p_m=(N-1)/2$ if $N$ is odd)}. This quantum number is associated to the spin of each irreducible block and defined as
\be
p=\frac{N}{2}-j_{\alpha_i} \, .
\ee
This means that $p=0$ corresponds to the block with the highest spin $J=N/2$ ({\em i.e.} the one with the lowest energy), $p=1$ corresponds to the first excited states corresponding to the next block(s), and so on. In terms of this quantum number and by correcting the eigenvalues in the form~\eqref{corrE}, the ferromagnetic energy spectrum without magnetic field is simply 
\be
E'(p,N)=\frac{I}{2}(-p^2 + (N+1)p) \, .
\ee
The energy levels~\eqref{corrE} for $N=2,3,4$ in the antiferromagnetic case, together with their degeneracies,  are shown in Figs.~\ref{qkskj1/2eigenB}, respectively. As expected, the presence of an external magnetic field partially breaks these degeneracies.

Therefore, the partition function for the spin-$1/2$ KS model is given by 
\begin{equation}
    Z_{N}=\sum_{i=1}^{k} \sum_{m_{\alpha_i}=-j_{\alpha_i}}^{m_{\alpha_i}=j_{\alpha_i}}  \text{exp}(-\beta E'(j_{\alpha_i},m_{\alpha_i},3 N/ 4)),
\end{equation}
where $\beta=1/k_{B}T$ and $k_{B}$ is the Boltzmann constant. As we have mentioned, irreducible blocks with the same spin $j_{\alpha_i}$ are obtained and, by taking into account the corresponding multiplicities $ d_{J N  \frac12}$~\eqref{degeneracies}, the previous formula can be rewritten as
\begin{equation}
    Z_{N}=\sum_{J}\sum_{m_J=-J}^{m_J=+J} d_{J N \frac12} \, \text{exp}(-\beta E'(J,m_J,3 N/4)),
    \label{partition1}
\end{equation}
where $J$ denotes all the different spins obtained in the block-diagonal decomposition, and the coefficients $d_{JN\frac12}$ represent the total multiplicity of a given energy level with spin $J$. 

The density of energy states can be obtained from the energy spectrum~\eqref{eigenvaluesKS} and the explicit expression of the degeneracies~\eqref{degeneracies}. Figure~\ref{kskj1/2Densidad} shows the density of states for $N=20,100,1000$ spins in the antiferromagnetic case (without an external magnetic field). {The plots corresponding to the ferromagnetic case are not included, as they coincide with those of the antiferromagnetic case under energy inversion.} As noted in~\cite{czachor2008energy}, in the ferromagnetic case, the ground and first excited energy states are the most separated energy levels and exhibit the lowest degeneracies. In contrast, in the antiferromagnetic case, this situation is reversed, and the lowest energy levels are the closest ones and also have the highest degeneracies. Moreover, in~\cite{van1993heisenberg} an integral formula for the partition function was obtained. 

The partition function~\eqref{partition1} can be now used to obtain the Helmholtz free energy as 
\be
F=-(k_{B}T/N)\log(Z_{N})\, ,
\ee 
and all the thermodynamic functions can be obtained from $F$. In particular, the specific heat, magnetic susceptibility, and magnetization are given by
\be 
    C_{V}=-T\left.\frac{\partial^{2}F}{\partial T^{2}}\right|_{h=0},\qquad
    \chi=-\left.\frac{\partial^{2}F}{\partial h^{2}}\right|_{h=0},\qquad
    M=-\frac{\partial F}{\partial h} \, .
    \label{eq:thermody}
\ee
In addition, as shown in~\cite{kittel1965development}, the Curie temperature corresponding to the ferromagnetic phase transition of the model can be estimated analytically, and it was proven that the phase transition develops very slowly as the number $N$ of spins increases.

\begin{figure}[p]
\centering
  \begin{minipage}{0.2\textwidth}
    \includegraphics[scale=0.7]{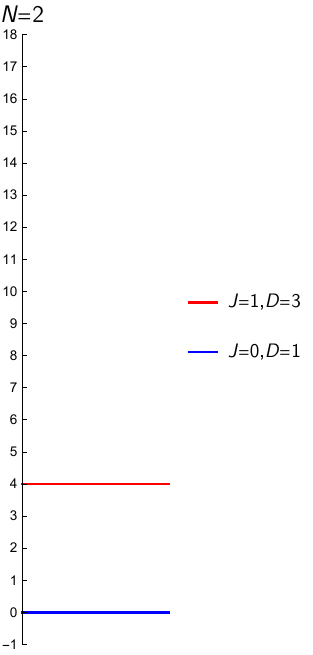}
  \end{minipage}
  \hspace{1.5mm}
  \begin{minipage}{0.2\textwidth}
    \includegraphics[scale=0.7]{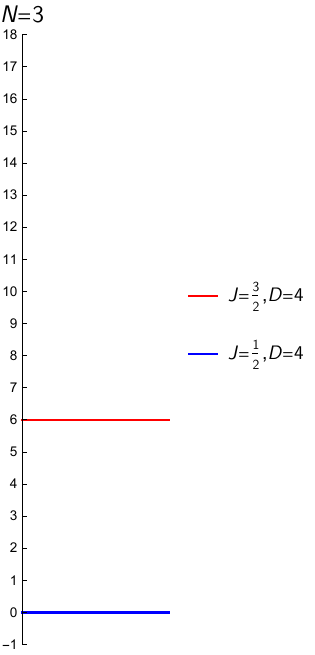}
  \end{minipage}
   \hspace{1mm}
  \begin{minipage}{0.2\textwidth}
    \includegraphics[scale=0.7]{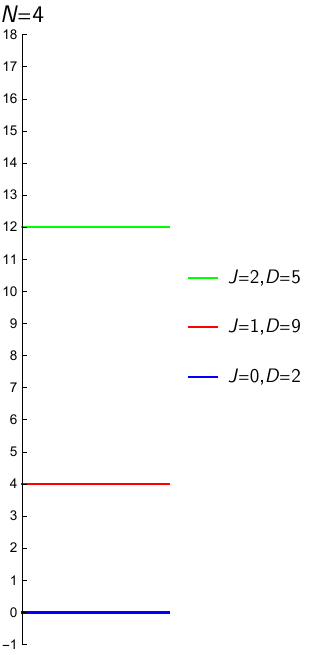}
  \end{minipage}
  \hspace{1mm}
  \begin{minipage}{0.2\textwidth}
    \includegraphics[scale=0.7]{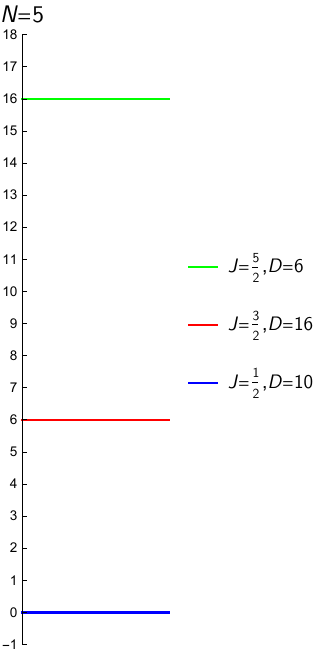}(a)
  \end{minipage}

\centering
  \begin{minipage}{0.2\textwidth}
    \includegraphics[scale=0.7]{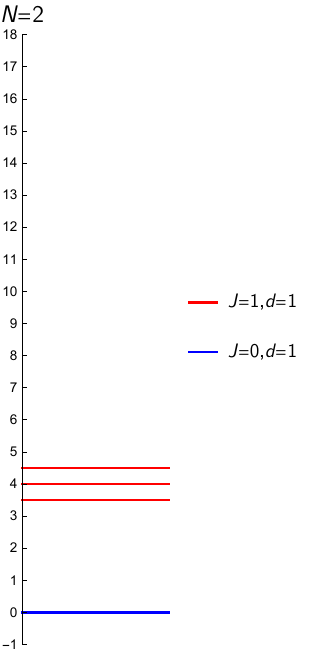}
  \end{minipage}
  \hspace{1.1mm}
  \begin{minipage}{0.2\textwidth}
    \includegraphics[scale=0.7]{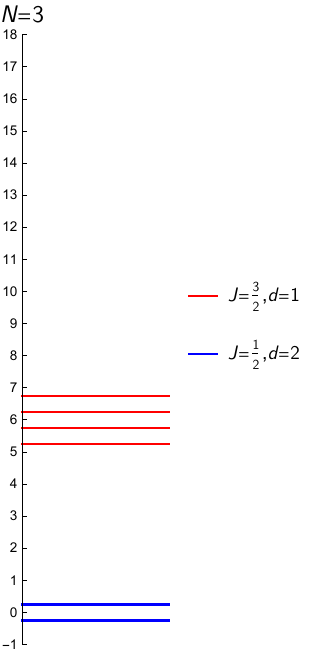}
  \end{minipage}
   \hspace{1mm}
  \begin{minipage}{0.2\textwidth}
    \includegraphics[scale=0.7]{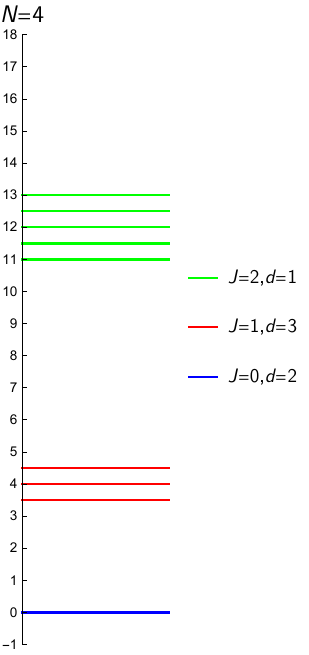}
  \end{minipage}
  \hspace{1mm}
  \begin{minipage}{0.2\textwidth}
    \includegraphics[scale=0.7]{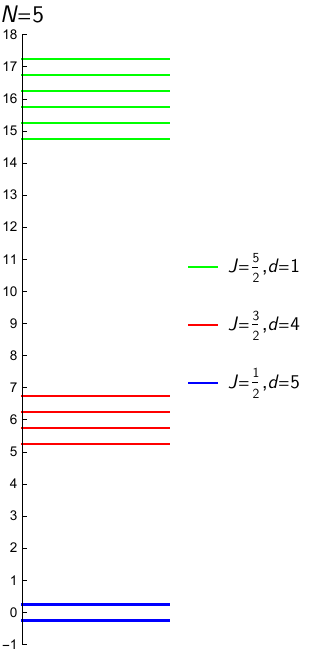}(b)
  \end{minipage}
  \caption{Energy levels (in the units $|I|/4$) and their degeneracies for $N=2,3,4,5$ spins with $j=1/2$ coupled antiferromagnetically (a) without an external magnetic field $(h=0)$ and (b) with an external magnetic field $(h=0.5)$. Adapted from~\cite{al1998exact}. {Note that $(1/2)^{\otimes 2} = 1 \oplus 0$, $(1/2)^{\otimes 3} = 3/2 \oplus (1/2)^{\oplus 2}$, $(1/2)^{\otimes 4} = 2 \oplus 1^{\oplus 3} \oplus 0^{2}$ and $(1/2)^{\otimes 5} = 5/2 \oplus (3/2)^{\oplus 4} \oplus (1/2)^{\oplus 4}$.}}
  \label{qkskj1/2eigenB}
\end{figure}

\begin{figure}[p]
    \centering
    \includegraphics[scale=0.9]{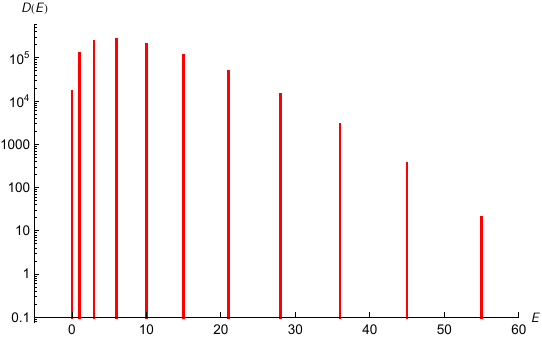} 
   \caption{Density of energy levels (with zero energy ground state) for $N=20$ in the antiferromagnetic case.}
  \label{kskj1/2Densidad}
\end{figure}

\begin{figure}[p]
    \centering
    \includegraphics[scale=0.9]{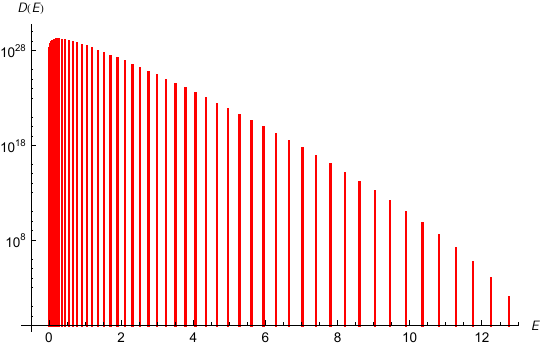}
  \caption{Density of energy levels for $N=100$, $I\rightarrow\frac{I}{N}$ in the antiferromagnetic case.}
\end{figure}

\begin{figure}[p]
  \centering
  \includegraphics[scale=0.9]{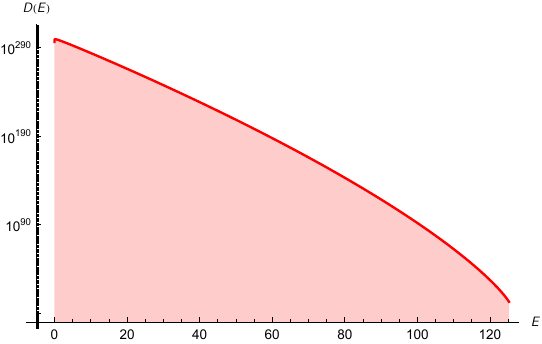}
  \caption{Density of energy levels for $N=1000$, $I\rightarrow\frac{I}{N}$ in the antiferromagnetic case.}
\end{figure}

\section{The $q$-deformed KS model}

As anticipated, the coalgebra symmetry of the KS model allows the construction of its $q$-deformation by making use of the quantum universal enveloping algebra {$ U_q(\mathfrak{su}(2))$}, where $q$ is a real parameter (hereafter referred to as quantum {$U_q(\mathfrak{su}(2))$} algebra), because the structure of its representation theory is exactly the same as that for {$ U(\mathfrak{su}(2))$} provided that $q$ is not a root of 1. In the following, we recall the basic notions of this quantum algebra and present the construction of the $q$-KS model by imposing its {$ U_q(\mathfrak{su}(2))$} coalgebra symmetry.

\subsection{The quantum algebra ${su}_q(2)$}

The three generators of the $U_q(\mathfrak{su}(2))$ algebra satisfy the following commutation relations, where $q\in  {\rm I\!R}^+$ (see~\cite{biedenharn1995quantum} and references therein):
\begin{equation}
[L_z,L_\pm]= \pm L_\pm,\qquad [L_+,L_-]= \left[ 2 L_z\right]_q = \frac{q^{L_z}-q^{-L_z}}{q^{1/2}-q^{-1/2}}  ,
\label{suq2}
\end{equation}
where we have made use of the $q$-symbol
\begin{equation}
[n]_q\coloneqq  \frac{q^{n/2}-q^{-n/2}}{q^{1/2}-q^{-1/2}}.
\label{eq:qnumber}
\end{equation}
We will frequently write $q=e^\eta$, which implies that
\be
 \left[ 2 L_z\right]_q = \frac{q^{L_z}-q^{-L_z}}{q^{1/2}-q^{-1/2}}= \frac{e^{\eta\,L_z}-e^{-\eta\,L_z}}{e^{\eta/2}-e^{-\eta/2}}=
  2 L_z +o[\eta^2],
\ee
and we recover the $U(\mathfrak{su}(2))$ Lie algebra in the ${q\rightarrow1\, (\eta\rightarrow 0)}$ limit. We remark that as a consequence of~\eqref{suq2}, the following relations hold:
\begin{equation}
q^{L_z} L_\pm q^{-L_z}=q^{\pm 1} L_\pm .
\label{qpm1}
\end{equation}

The coalgebra structure for the $U_q(\mathfrak{su}(2))$ algebra is generated by the {coproduct map}
\begin{equation}
\Delta (L_\pm)= q^{-L_z/2}\otimes L_\pm+ L_\pm \otimes q^{L_z/2},\qquad \Delta (L_z)= \mathbf{I} \otimes L_z+ L_z \otimes \mathbf{I} ,
\label{qcop}
\end{equation}
which is an algebraic homomorphism with respect to~\eqref{suq2} and leads to the undeformed coproduct $\co_0$~\eqref{coprim} in the limit $q\rightarrow1$ ($\eta\rightarrow 0$). The generalisation to an arbitrary number $N$ of tensor products is given by
{
\bea
\co^{(N)}(L_\pm)=L_\pm\otimes (q^{L_z/2})^{\otimes(N-1)} + q^{-L_z/2}\otimes L_\pm\otimes (q^{L_z/2})^{\otimes(N-2)}  +
\dots  + (q^{L_z/2})^{\otimes(N-1)}\otimes L_\pm ,
\label{coNq}
\eea}
and for the $L_z$ generator, the deformed coproduct map coincides with the undeformed one, $\co^{(N)}(L_z)\equiv \co^{(N)}_0(L_z)$~\eqref{coN}. 

Irreducible representations for $U_q(\mathfrak{su}(2))$ are defined through the action of the algebra generators onto an state $|j ,m\rangle$ in the form
\begin{equation}
L_\pm |j, m\rangle\, :=\, \sqrt{[j\mp m]_q [j\pm m+1]_q}|j ,m\pm 1\rangle,\qquad  L_z |j, m\rangle=m |j, m\rangle.
\label{repsuq2}
\end{equation}
Moreover, the deformed Casimir operator for the algebra~\eqref{suq2} reads
\begin{equation}
C_q= L_- L_+ + [L_z]_q[L_z+\mathbf{I}]_q=L_+ L_- + [L_z]_q[L_z-\mathbf{I}]_q \, ,
\label{qcas}
\end{equation}
and the eigenvalues of $C_q$ in the representation states $ |j, m\rangle$ are just $[j]_q[j+1]_q$. 
Therefore, if $q$ is not a root of unity, the representation theory of $U_q(\mathfrak{su}(2))$ is just structurally the same as that of $U(\mathfrak{su}(2))$ (see~\cite{biedenharn1995quantum}). 

Moreover, the tensor product of two $U_q(\mathfrak{su}(2))$ irreducible representations $j_1$ and $j_2$ is constructed using the deformed coproduct $\co$, and its decomposition into irreducible modules with spin $j$  is the same as in the undeformed case, as we obtain invariant subspaces with $j=(j_1+j_2),\dots, | j_1-j_2 |$. This result can obviously be generalised to the direct sum of $N$ irreducible representations of $U(\mathfrak{su}(2))$ and constitutes the keystone for the exact solvability of the $q$-deformed KS model introduced in the following.

\subsection{The generic $q$-Kittel Shore model}

The generic $q$-KS Hamiltonian with a external magnetic field is defined by substituting the undeformed coproducts and Casimir operators in~\eqref{KScoalgebrah0} by their $q$-deformed counterparts, namely
\begin{equation}
    \tilde H_{KS}^q=-\frac{I}{2}\left(\Delta^{(N)}(C_q)-\sum_{i=1}^N C_q^{(i)}\right) -\gamma h \Delta^{(N)}(L_{z})\, ,
\label{qKScoalgebrah0}
\end{equation}
where the $q$-deformed Casimir operator $C_q$ is the $U_q(\mathfrak{su}(2))$ Casimir operator given by~\eqref{qcas}. Again, the deformed coproduct map $\Delta^{(N)}$ is an algebra homomorphism with respect to the defining relations of the $U_q(\mathfrak{su}(2))$ algebra~\eqref{suq2}
\be
\conm{\Delta^{(N)}(C_q)}{\Delta^{(N)}(X)}=\Delta^{(N)}(\conm{C_q}{X})=0\, ,
\qquad \forall\,  X \in U_q(\mathfrak{su}(2)).
\label{qschur}
\ee
Therefore, by defining the $q$-KS Hamiltonian in the form~\eqref{qKScoalgebrah0}, we have preserved, by construction, the $U_q(\mathfrak{su}(2))$ coalgebra symmetry of the model. This implies that the Hamiltonian~\eqref{qKScoalgebrah0} commutes with $\Delta^{(N)}(L_{z})$, which is also the total $z$-component of the $q$-deformed spin operators. Moreover, in the $h=0$ case, the collective raising and lowering $q$-spin operators~\eqref{coNq} also provide the commuting operators with $H_{KS}^q$. Finally, notice that the Hamiltonian~\eqref{qKScoalgebrah0} is indeed Hermitian, as can be straightforwardly verified.

Consequently, we can mimic the construction performed for the $U(\mathfrak{su}(2))$ algebra. We start from a block-diagonal basis~\eqref{blockdbasis} for the tensor product of $N$ irreducible representations of $U_q(\mathfrak{su}(2))$, where the operator $C_q^{(N)}\equiv \Delta^{(N)}(C_q)$ arising from the $q$-deformed coproduct and Casimir  will again be represented as a block-diagonal matrix. Within each  irreducible block defined by a set of basis vectors~\eqref{basisbd}, the relations~\eqref{qschur} imply that $C_q^{(N)}$ has to be proportional to the identity operator with eigenvalues
\be
\Delta^{(N)}(C)\,|j_{\alpha_i}, m_{\alpha_i}\rangle = [j_{\alpha_i}]_q\, [j_{\alpha_i}+1]_q\, |j_{\alpha_i}, m_{\alpha_i}\rangle ,
\qquad
\forall \alpha_i=1,\dots,k
 ,
\qquad
m_{\alpha_i}=-j_{\alpha_i},\dots,j_{\alpha_i}  .
\ee
Also, the operator $\sum_{i=1}^N C_q^{(i)}$ acting on the tensor product basis~\eqref{tpbasis} is just the identity times the factor $K_q=\sum_{i=1}^N [j_i]_q\,[(j_i+1)]_q$,  and the eigenvectors of the $q$-KS Hamiltonian~\eqref{qKScoalgebrah0} are again the block-diagonal basis vectors~\eqref{basisbd} with eigenvalues given by
\be
H_{KS}^q |j_{\alpha_i}, m_{\alpha_i}\rangle=-\frac{I}{2}\left(\Delta^{(N)}(C_q)-\sum_{i=1}^N C_q^{(i)}\right)\,|j_{\alpha_i}, m_{\alpha_i}\rangle =
-\frac{I}{2}\left( [j_{\alpha_i}]_q\,[j_{\alpha_i}+1]_q-K_q\right)\,|j_{\alpha_i}, m_{\alpha_i}\rangle .
\ee
Finally, when $h\neq 0$, since the coproduct for the $L_z$ generator is not deformed, by following the same reasoning as in the non-deformed case we have that
\be
-\gamma h \Delta^{(N)}(L_{z})  |j_{\alpha_i}, m_{\alpha_i}\rangle =-\gamma\,h\, m_{\alpha_i} |j_{\alpha_i}, m_{\alpha_i}\rangle  .
\ee

In conclusion, the eigenvalues of the generic $q$-KS model for $N$ arbitrary spins are 
\be
\tilde E_q(j_{\alpha_i},m_{\alpha_i},K_q)=-\frac{I}{2}\left( [j_{\alpha_i}]_q \, [j_{\alpha_i}+1]_q-K_q\right) -\gamma\,h\, m_{\alpha_i}.
\label{eigenvaluesqKS}
\ee
Note that when comparing this expression with~\eqref{eigenvaluesqKS}, all numbers are substituted by $q$-numbers, except for the magnetic-field term. In addition, the ground-state eigenvalue can be set to zero for any $N$ by subtracting the ground-state energy from all the eigenvalues.

Let us make the abstract definition~\eqref{qKScoalgebrah0} of the $q$-KS model more explicit in terms of the $U_q(\mathfrak{su}(2))$ generators. A first step consists in rewriting it in the form~\eqref{qKScoalgebra}, namely
\begin{equation}
    \tilde H_{KS}^q=-\frac{I}{2} \left(\Delta^{(N)} (L_{-})\Delta^{(N)} (L_{+})+
    \left[ \Delta^{(N)} (L_{z})\right]_q \left[\Delta^{(N)} (L_{z})+\Delta^{(N)}(\mathbf{I})\right]_q
    -\sum_{i=1}^N C_q^{(i)}\right)-\gamma h \Delta^{(N)}(L_{z}) .
\label{qKScoalgebra}
\end{equation}

Now we consider the explicit form of the $N$-th deformed coproducts~\eqref{coNq}
 \begin{equation}
\Delta^{(N)}(L_{\pm})=  \sum_{i=1}^N \exp\left[-\frac{\eta}{2}\sum_{j=1}^{i-1}L_z^{(j)}\right]L_{\pm}^{(i)} \exp\left[\frac{\eta}{2} \sum_{h=i+1}^{N}L_z^{(h)}\right], \qquad \Delta^{(N)}(L_{z})=  \sum_{i=1}^N  L_z^{(i)},
\label{eq:coproduct_n}
\end{equation}
where the superindex $i$ represents the spin for the $i$-th particle and we have used that $q=e^\eta$ (which, as we will see, will be useful in the following). 
Expanding the Hamiltonian~\eqref{qKScoalgebra}, we find the expression
\begin{align}
\!\!\!\!\!\!\!\!\!\!    
\tilde H_{KS}^{q}=&-\frac{I}{2}\left(\sum_{i=1}^{N}\text{ exp} \left[-\eta \sum_{j=1}^{i-1}L_{z}^{(j)}\right]\left(C^{(i)}_{q}-[L_{z}^{(i)}]_{q}\left[L_{z}^{(i)}+\mathbf{I}^{(i)}\right]_{q}\right)\text{ exp} \left[\eta\sum_{h=i+1}^{N} L_{z}^{(h)}\right]+\sum_{i=1}^{N-1}\sum_{k=i+1}^{N}\prod_{t=1}^{i-1}\text{ exp} \left[-\eta L_{z}^{(t)}\right]\cdot\right. \notag  \\ 
    &\cdot\text{ exp}\left[-\eta \frac{L_{z}^{(i)}}{2}\right]\left(L_{+}^{(i)}L_{-}^{(k)}+L_{-}^{(i)}L_{+}^{(k)}\right)\text{ exp}\left[\eta \frac{L_{z}^{(k)}}{2}\right]\prod_{r=k+1}^{N}\text{ exp} \left[\eta L_{z}^{(r)}\right]+\left[\sum_{i=1}^{N}L_{z}^{(i)}\right]_{q}\left[\sum_{s=1}^{N}L_{z}^{(s)}+\mathbf{I}^{(N)}\right]_{q}- \notag \\       
    &-\sum_{i=1}^N C_q^{(i)}\Biggl)-\gamma h \sum_{i=1}^{N}L_{z}^{(i)} .
\end{align}

Now, by taking into account that $
q^{L_z/2} L_\pm q^{-L_z/2}=q^{\pm 1/2} L_\pm 
$, we obtain the final expression
\begin{align}
    \tilde H_{KS}^{q}=&-\frac{I}{2}\left(\sum_{i=1}^{N}\text{ exp} \left[-\eta \sum_{j=1}^{i-1}L_{z}^{(j)}\right]L_{-}^{(i)}L_{+}^{(i)}\text{ exp} \left[\eta\sum_{h=i+1}^{N} L_{z}^{(h)}\right]+\sum_{i=1}^{N-1}\sum_{r=i+1}^{N}\left(e^{\eta/2}L_{-}^{(i)}L_{+}^{(r)}+e^{-\eta/2}L_{+}^{(i)}L_{-}^{(r)}\right)\cdot\right. \notag  \\ 
    &\cdot\text{ exp}\left[-\eta \frac{L_{z}^{(i)}}{2}\right]\cdot\text{ exp} \left[\eta \frac{L_{z}^{(r)}}{2}\right]\cdot\prod_{t=1}^{i-1}\text{ exp} \left[-\eta L_{z}^{(t)}\right]\prod_{k=r+1}^{N}\text{ exp} \left[\eta L_{z}^{(k)}\right]+\left[\sum_{i=1}^{N}L_{z}^{(i)}\right]_{q}\left[\sum_{s=1}^{N}L_{z}^{(s)}+\mathbf{I}^{(N)}\right]_{q}- \notag \\       
    &-\sum_{i=1}^N C_q^{(i)}\Biggl)-\gamma h \sum_{i=1}^{N}L_{z}^{(i)} ,
    \label{finalqKSpm}
\end{align}
which, for $q=1$ (equivalently, $\eta=0$), leads to the undeformed KS Hamiltonian~\eqref{KSco}. 

Note that the $q$-KS Hamiltonian can be also written in terms of the $L_x$, $L_y$, and $L_z$ spin operators,
where we consider the same change of basis as in the non-deformed case, namely
\begin{equation}
    L_x = \frac{1}{2} (L_+ + L_-), \qquad L_y = - \frac{i}{2} (L_+ - L_-)\, .
\end{equation}
In this way, we have that
\begin{equation}
    \tilde H^q_{KS}=-\frac{I}{2} \left(\Delta (L_{x})^{2}+\Delta (L_{y})^{2}-\frac{1}{2} [2\Delta (L_{z})]_{q}+[\Delta (L_{z})]_{q}[\Delta (L_{z})+\Delta (\mathbf{I})]_{q}-\sum_{i=1}^N C_q^{(i)}\right)-\gamma h \Delta^{(N)}(L_{z}) ,
    \label{eq:qhamiltonianxyz}
\end{equation}
which leads to the final expression
\begin{align}
    \tilde H_{KS}^{q(N)}=&-\frac{I}{2}\left(\sum_{i=1}^{N}\text{ exp} \left[-\eta \sum_{j=1}^{i-1}L_{z}^{(j)}\right]\left((L_{x}^{(i)})^{2}+(L_{y}^{(i)})^{2}-\frac{1}{2}[2L_{z}^{(i)}]_{q}\right)\text{ exp} \left[\eta\sum_{h=i+1}^{N} L_{z}^{(h)}\right]+\sum_{i=1}^{N-1}\sum_{l=i+1}^{N}\left[\left(L_{x}^{(i)}L_{x}^{(l)}+\right.\right.\right. \notag  \\ 
    &\left.\left.\left.+L_{y}^{(i)}L_{y}^{(l)}\right)\left(e^{\eta/2}+e^{-\eta/2}\right)+i\left(L_{x}^{(i)}L_{y}^{(l)}-L_{y}^{(i)}L_{x}^{(l)}\right)\left(e^{\eta/2}-e^{-\eta/2}\right)\right]\text{ exp}\left[-\eta \frac{L_{z}^{(i)}}{2}\right]\cdot\text{ exp} \left[\eta \frac{L_{z}^{(l)}}{2}\right]\cdot\right. \notag \\       
    &\left.\cdot\prod_{t=1}^{i-1}\text{ exp} \left[-\eta L_{z}^{(t)}\right]\prod_{k=l+1}^{N}\text{ exp} \left[\eta L_{z}^{(k)}\right]+\left[\sum_{i=1}^{N}L_{z}^{(i)}\right]_{q}\left[\sum_{s=1}^{N}L_{z}^{(s)}+\mathbf{I}^{(N)}\right]_{q}-\sum_{i=1}^N C_q^{(i)}\right)-\gamma h \sum_{i=1}^{N}L_{z}^{(i)}.
\label{finalqKS}
\end{align}

As can be seen from the previous expressions, the explicit form of the deformation in terms of the quantum algebra generators is far from simple. Nevertheless, the shape of the deformation is absolutely dictated by the $U_q(\mathfrak{su}(2))$ symmetry to preserve all the integrability properties of the KS model and, despite its complexity, leads to the simple form~\eqref{eigenvaluesqKS} for its eigenvalues, which preserves the same degeneracies as in the undeformed model but becomes nonlinearly modified in terms of the quantum deformation parameter $\eta$.

\section{The spin-$1/2$ $q$-KS model}

\subsection{Eigenvectors and eigenvalues}

Taking into account that the 2-dimensional $j=1/2$ fundamental irreducible representation~\eqref{repsuq2} for $U_q(\mathfrak{su}(2))$ is given by the undeformed Pauli matrices~\eqref{paulipm}, this means that the $q$-KS model for $N$ copies of the $U_q(\mathfrak{su}(2))$ algebra gives rise to a spin Hamiltonian whose explicit form is obtained by substituting 
\begin{equation}
L_{\pm}^{(i)} = 
\sigma_{\pm}^{(i)} , \qquad
L_z^{(i)} = 
\frac12\sigma_z^{(i)} ,
\qquad
i=1,\dots, N,
\label{spin12}
\end{equation}
within~\eqref{finalqKSpm}, where $e^\eta=q$, namely 
\begin{align}
    \tilde H_{KS}^{q (N)}=&-\frac{I}{2} \Biggl(\,\sum_{i=1}^{N}\text{ exp} \left[-\eta \sum_{j=1}^{i-1}\frac12 \sigma_{z}^{(j)}\right]\sigma_{-}^{(i)}\sigma_{+}^{(i)}\text{ exp} \left[\eta\sum_{h=i+1}^{N} \frac12 \sigma_{z}^{(h)}\right] \notag \\
    &
    +\sum_{i=1}^{N-1}\sum_{r=i+1}^{N}\left(e^{\eta/2}\sigma_{-}^{(i)}\sigma_{+}^{(r)}+e^{-\eta/2}\sigma_{+}^{(i)}\sigma_{-}^{(r)}\right)\cdot \notag  \\ 
    &\cdot\text{ exp}\left[-\eta \frac{\sigma_{z}^{(i)}}{4}\right]\cdot\text{ exp} \left[\eta \frac{\sigma_{z}^{(r)}}{4}\right]\cdot\prod_{t=1}^{i-1}\text{ exp} \left[-\frac \eta 2 \sigma_{z}^{(t)}\right]\prod_{k=r+1}^{N}\text{ exp} \left[\frac \eta 2 \sigma_{z}^{(k)}\right] \notag \\       
    &+\left[\sum_{i=1}^{N}\frac12 \sigma_{z}^{(i)}\right]_{q}\left[\sum_{s=1}^{N} \frac12 \sigma_{z}^{(s)}+\mathbf{I}^{(N)}\right]_{q}- N\,[1/2]_q\,[3/2]_q\Biggl)-\gamma h \sum_{i=1}^{N}\frac12 \sigma_{z}^{(i)} \,.
    \label{finalqKSpmpauli}
\end{align}

We stress that this is a new integrable spin model with long-range interactions that, in contrast to the undeformed KS model, presents non-identical couplings among  different spins, which are determined by the deformed coproducts. Nevertheless, the underlying $U_q(\mathfrak{su}(2))$ coalgebra symmetry of the model makes that these involved interactions are such that the eigenvalues of this model can be straightforwardly deduced from~\eqref{eigenvaluesqKS} and read
\be
\tilde E_q(j_{\alpha_i},m_{\alpha_i},K_q)=-\frac{I}{2}\left( [j_{\alpha_i}]_q \, [j_{\alpha_i}+1]_q-
N\,[1/2]_q\,[3/2]_q
\right) -\gamma\,h\, m_{\alpha_i} ,
\label{eigenvaluesqKS12}
\ee
which should be compared to \eqref{eigenvaluesKS}. On the other hand, eigenvectors for $N=2,3,4,\dots$ can be obtained by applying in a recurrent way the well-known expressions for the $q$-Clebsch--Gordan coefficients (see~\cite{alvarez2024analog} and references therein)
\begin{equation}
\begin{split}
CG[j_1,m_1,j_2,m_2,j,m]\,=&\,\delta _{m,m_1+m_2} q^{\frac{1}{2} (j_1 m_2-j_2 m_1)-\frac{1}{4} (-j+j_1+j_2) (j+j_1+j_2+1)}
\\
 &\sqrt{\frac{[2 j+1]_q  [j+m]_q! [j_2-m_2]_q![j+j_1-j_2]_q! [-j+j_1+j_2]_q! [j+j_1+j_2+1]_q! }{[j-m]_q! [j_1-m_1]_q!  [j_1+m_1]_q! [j_2+m_2]_q! [j-j_1+j_2]_q!  }} 
\\
&\sum _{n=0}^{\min (-j+j_1+j_2,j_2-m_2)} \frac{ [2 j_2-n]_q!  (-1)^{-j+j_1+j_2+n} q^{\frac{1}{2} n (j_1+m_1)} [j_1+j_2-m-n]_q! }{ [n]_q!  [j_2-m_2-n]_q!   [-j+j_1+j_2-n]_q!  [j+j_1+j_2-n+1]_q! } \, .
\end{split}
\label{eq:qcg}
\end{equation}

As a consequence, the partition function $Z_{N}^q$ of this model with $h=0$ is directly obtained from~\eqref{partition1}, and reads
\begin{equation}
    Z_{N}^q=\sum_{J}\sum_{m_J=-J}^{m_J=+J} d_{NJ\tfrac12} \, \text{exp}(-\beta E'_q(J,m_J,K_q)),
    \label{partition}
\end{equation}
where $E'_q(J,m_J,K_q)$ is obtained from~\eqref{eigenvaluesqKS12} with $h=0$ and the degeneracies $d_{NJ\tfrac12}$ are the same as those in the undeformed model~\eqref{degeneracies} because of the identical structure of the representation theory of $U_q(\mathfrak{su}(2))$ and $U(\mathfrak{su}(2))$. From this expression, all the thermodynamic properties of the $q$-deformed model can be derived and studied. 
As in the undeformed case, the energy levels $E'_q$ are obtained by subtracting the energy of the ground state.

Regarding the density of the (corrected) energy levels of the $q$-deformed model, we can easily see that the deformation enlarges the distance between energy states, as shown in Figure~\ref{kskj1/2DensityD}, which plots the  densities of the energy levels for the deformed model with $N=20,100,1000$ and different values of $\eta$, because the larger $N$ is, the smaller $\eta$ must be considered in order to obtain reasonable values for the energies. {Again, in the deformed model, the energy spectra for the ferromagnetic and antiferromagnetic couplings remain fully symmetric under energy inversion. Consequently, only the antiferromagnetic case is displayed.} 
 
\begin{figure}[p]
  \centering
  \includegraphics[scale=0.9]{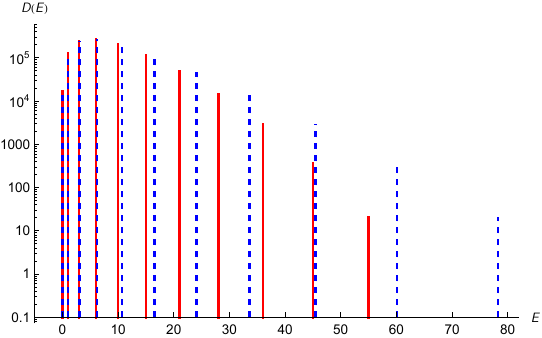}
  \caption{Density of energy levels for $N=20$ as a function of the parameter $\eta=0$ (red) and $\eta=0.2$ (dashed blue) in the antiferromagnetic case.}
  \label{kskj1/2DensityD}
\end{figure}

\begin{figure}[p]
  \centering 
  \includegraphics[scale=0.9]{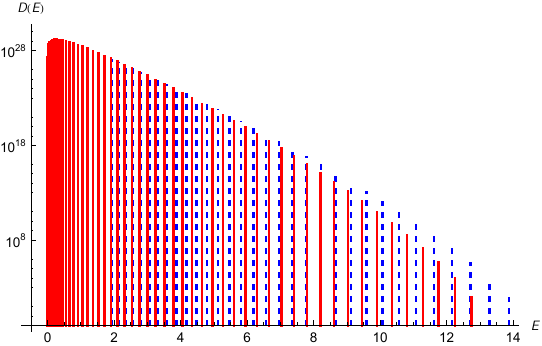}
  \caption{Density of energy levels for $N=100$ as a function of the parameter $\eta=0$ (red) and $\eta=0.02$ (dashed blue) in the antiferromagnetic case.}
\end{figure}

\begin{figure}[p]
  \centering
  \includegraphics[scale=0.9]{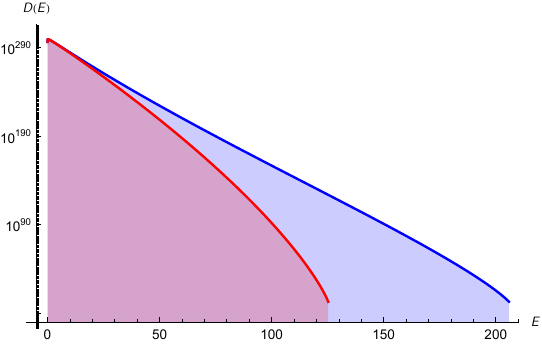}
  \caption{Density of energy levels for $N=1000$ as a function of the parameter $\eta=0$ (red) and $\eta=0.005$ (blue) in the antiferromagnetic case.}
\end{figure}

\subsection{The $N=2$ and $3$ cases}

To illustrate the main features of the spin-$1/2$ $q$-KS model, an explicit study of the $N=2,3,4$ cases is required.  The eigenvalues of the $q$-KS antiferromagnetic Hamiltonian in the presence of an external magnetic field are plotted in Figure~\ref{qkskj1/2eigenD}. Thus, the nonlinear effect of the deformation in the energy gaps between consecutive eigenvalues of the model can be clearly observed. The degeneracies for each energy level are not modified with respect to the undeformed KS Hamiltonian, and the addition of a constant external magnetic field would generate the same uniform Zeeman splittings as in the $q=1$ case. Indeed, the antiferromagnetic and ferromagnetic spectra are inverted versions of the very same energy level structure.

The eigenvectors for the $N=2,3,4$ $q$-KS model can be computed using the $q$-Clebsch--Gordan coefficients~\eqref{eq:qcg}, which are presented in the following. In this way, the role that the $q$-Dicke states~\cite{li2015entanglement, raveh2024q} play in the model is shown, which provides a basis for the fundamental eigenspace in the ferromagnetic case and for the most excited state in the antiferromagnetic case. The remaining eigenstates are no longer $q$-Dicke eigenstates and are explicitly given and analysed.

\subsubsection{The $N=2$ case and the $q$-exchange operator}

In this case the $q$-KS Hamiltonian of two interacting spins without magnetic field is given, in the fundamental representation~\eqref{spin12}, by 
\begin{align}
\! \! \! \! \! \! \! \! \! \! \! \! 
   H_{KS}^{q(2)}=&-\frac{I}{2} \Biggl(
    \sigma_{-}^{(1)}\sigma_{+}^{(1)}\text{ exp} \left[\eta  \frac12 \sigma_{z}^{(2)}  \right] +
      \text{ exp} \left[-\eta \frac12 \sigma_{z}^{(1)}\right]\sigma_{-}^{(2)}\sigma_{+}^{(2)} 
     + \left(e^{\eta/2}\sigma_{-}^{(1)}\sigma_{+}^{(2)}+e^{-\eta/2}\sigma_{+}^{(1)}\sigma_{-}^{(2)} \right) \cdot \notag \\ 
   &\cdot\text{ exp}\left[-\eta \frac{\sigma_{z}^{(1)}}{4}\right]\cdot\text{ exp} \left[\eta \frac{\sigma_{z}^{(2)}}{4}\right]+\left[\sum_{i=1}^{2}\frac12 \sigma_{z}^{(i)}\right]_{q}\left[\sum_{s=1}^{2} \frac12 \sigma_{z}^{(s)}+\mathbf{I}^{(2)}\right]_{q} 
   -2\,[1/2]_q\,[3/2]_q\Biggl) \, .
    \label{finalqKSpmpauliN2q}
\end{align}
Obviously, the $\eta\to 0$ limit of this expression is the $N=2$ KS model rewritten in the form
\begin{align}
    H_{KS}^{(2)}& =-\frac{I}{2} \Biggl( \left(
    \sigma_{-}^{(1)}\sigma_{+}^{(1)} +
   \sigma_{-}^{(2)}\sigma_{+}^{(2)} 
     + \sigma_{-}^{(1)}\sigma_{+}^{(2)}+\sigma_{+}^{(1)}\sigma_{-}^{(2)} \right)
   +\left(\sum_{i=1}^{2}\frac12 \sigma_{z}^{(i)}\right)^2 + \left(\sum_{s=1}^{2} \frac12 \sigma_{z}^{(s)}\right) -3/2\Biggl) \notag \\       
    & 
    =- I\, \Vec{\sigma}_{1}\cdot\Vec{\sigma}_{2} \, .
    \label{exchangeop}
\end{align}
The structural complexity of the $q$-deformation can be appreciated by comparing the latter expression with~\eqref{finalqKSpmpauliN2q}. In fact, the operator~\eqref{finalqKSpmpauliN2q} can be taken as the definition of the $q$-exchange interaction induced by the deformed coalgebra structure of $U_q(\mathfrak{su}(2))$. In matrix form, the operator~\eqref{finalqKSpmpauliN2q} reads
\be
 H_{KS}^{q(2)}=-\frac{I}{\left(\sqrt{q}+1\right)^2}\left(\begin{pmatrix}
        \frac{q^2+3}{2\sqrt{q}} & 0 & 0 & 0 \\
        0 & \frac{1-q}{2\sqrt{q}} & \frac{1}{2} & 0 \\
        0 & \frac{1}{2} & \frac{(q-1) \sqrt{q}}{2} & 0  \\
        0 & 0 & 0 & \frac{q^2+3}{2 \sqrt{q}}  \\
    \end{pmatrix}-\begin{pmatrix}
        \frac{1}{\sqrt{q}} & 0 & 0 & 0 \\
        0 & q & 0 & 0 \\
        0 & 0 & 1 & 0  \\
        0 & 0 & 0 & \frac{1}{\sqrt{q}} \\
    \end{pmatrix}\right)\,,
    \label{qKSN2}
\ee
which is manifestly Hermitian, and  in the limit $q\to 1$ provides the usual form of the exchange operator~\eqref{exchangeop}:
\be
 H_{KS}^{q(2)}=-\frac{I}{2}\left(\begin{pmatrix}
       1 & 0 & 0 & 0 \\
        0 & 0 & 1 & 0 \\
        0 & 1 & 0 & 0  \\
        0 & 0 & 0 & 1  \\
    \end{pmatrix}- \frac{1}{2}\begin{pmatrix}
        1 & 0 & 0 & 0 \\
        0 & 1 & 0 & 0 \\
        0 & 0 & 1 & 0  \\
        0 & 0 & 0 & 1 \\
    \end{pmatrix}\right) \,.
\ee

In the ferromagnetic case, the ground state for~\eqref{finalqKSpmpauliN2q} with $h=0$  can be straightforwardly shown to be the $J=1$ subspace, which is generated by the $q$-triplet 
\begin{equation}
\ket{1,-1}= |\hspace{-0.1cm}\down\down\rangle,\qquad \ket{{1,0}}=  \frac{1}{\sqrt{[2]_{q}}} \left( q^{1/4} |\hspace{-0.1cm}\down\up\rangle +q^{-1/4}|\hspace{-0.1cm}\up\down\rangle\right),\qquad \ket{{1,1}}= |\hspace{-0.1cm}\up\up\rangle,
\label{eq:2qt}
\end{equation}
and the singlet with $j=0$ is the only excited state
\begin{equation}
\ket{{0,0}}= \frac{1}{\sqrt{[2]_{q}}} \left(- q^{-1/4} |\hspace{-0.1cm}\down\up\rangle +q^{1/4}|\hspace{-0.1cm}\up\down\rangle  \right) .
\label{eq:2qs}
\end{equation}
The situation is reversed in the antiferromagnetic case, and when the magnetic field $h\neq 0$ is included, the triplet becomes nondegenerate.

The states~\eqref{eq:2qt} are just the $q$-Dicke states for $j=1$. The states $|1, 1\rangle$ and $|1, -1\rangle$ are factorised, where $|1, 0\rangle$ and $|0, 0\rangle$ are $q$-Bell states whose entanglement is not maximal unless $q\to 1$ (the limit of the undeformed Bell states). However, the limit $q\to\infty$ gives rise to factorised states, thus showing that the $q$-deformation of the model (or, equivalently, the $q$-exchange operator~\eqref{finalqKSpmpauliN2q}) induces a loss of entanglement governed by the deformation parameter $q$.

{In Figure~\ref{qkskj1/2eigenD} we present the deformed version of Figure~\ref{qkskj1/2eigenB}, showing the energy levels of the model as a function of the deformation parameter $\eta=\log q$, including the Zeeman splitting. Note that although for a given $\eta$ the global shape of the spectrum is preserved, higher energy levels grow very rapidly in terms of $\eta$. As we will see in Section 6, this fact will be essential as far as the thermodynamical properties of the model are concerned.
}

\begin{figure}[H]
\centering
  \begin{minipage}{0.2\textwidth}
    \includegraphics[scale=0.7]{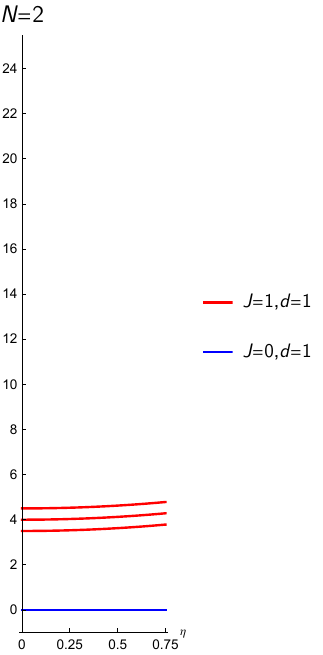}
  \end{minipage}
  \hspace{1mm}
  \begin{minipage}{0.2\textwidth}
    \includegraphics[scale=0.7]{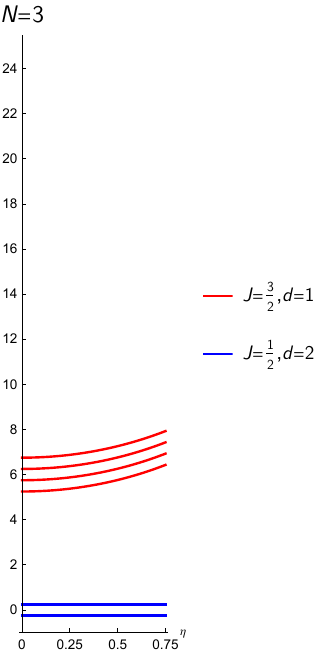}
  \end{minipage}
   \hspace{1mm}
  \begin{minipage}{0.2\textwidth}
    \includegraphics[scale=0.7]{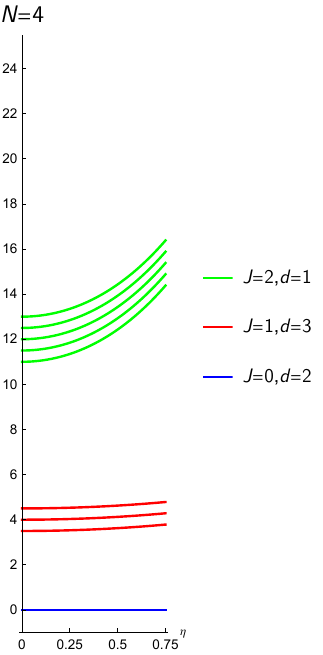}
  \end{minipage}
  \hspace{1mm}
  \begin{minipage}{0.2\textwidth}
    \includegraphics[scale=0.7]{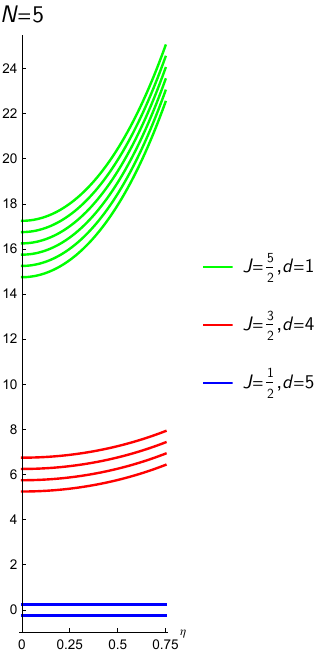}
  \end{minipage}
  \caption{Energy levels (in the units $|I|/4$) for $N=2,3,4,5$ spins $j=1/2$ respectively coupled antiferromagnetically as a function of the deformation parameter $\eta$ displaced from the fundamental level (a) without an external magnetic field $(h=0)$ and (b) with an external magnetic field $(h=0.5)$.}
  \label{qkskj1/2eigenD}
\end{figure}

%%%%%%%%%%%%%%%%%%%%%%%%%%%%%%%%%%%%%%%%%%%%%%

\subsubsection{The $N=3$ case}

When three spins are considered, the $q$-KS model in the fundamental representation~\eqref{spin12} and with $h=0$ becomes
\begin{align}
    H_{KS}^{q(3)}=&-\frac{I}{2} \Biggl(\,\sigma_{-}^{(1)}\sigma_{+}^{(1)}\text{ exp} \left[\eta\frac12( \sigma_{z}^{(2)}+\sigma_{z}^{(3)})\right]+\,\text{ exp} \left[-\eta\frac12\sigma_{z}^{(1)}\right]\sigma_{-}^{(2)}\sigma_{+}^{(2)}\text{ exp} \left[\eta\frac12\sigma_{z}^{(3)}\right]+ \notag \\
    &
    +\,\text{ exp} \left[-\eta\frac12(\sigma_{z}^{(1)}+\sigma_{z}^{(2)})\right]\sigma_{-}^{(3)}\sigma_{+}^{(3)}+\left(e^{\eta/2}\sigma_{-}^{(1)}\sigma_{+}^{(2)}+e^{-\eta/2}\sigma_{+}^{(1)}\sigma_{-}^{(2)}\right)\text{ exp}\left[-\eta \frac{\sigma_{z}^{(1)}}{4}\right]\cdot \notag  \\ 
    &\cdot\text{ exp} \left[\eta \frac{\sigma_{z}^{(2)}}{4}\right]\cdot\text{ exp} \left[\frac \eta 2 \sigma_{z}^{(3)}\right]+\left(e^{\eta/2}\sigma_{-}^{(1)}\sigma_{+}^{(3)}+e^{-\eta/2}\sigma_{+}^{(1)}\sigma_{-}^{(3)}\right)\text{ exp}\left[-\eta \frac{\sigma_{z}^{(1)}}{4}\right]\cdot \notag \\       
    &\cdot\text{ exp} \left[\eta \frac{\sigma_{z}^{(3)}}{4}\right]+\left(e^{\eta/2}\sigma_{-}^{(2)}\sigma_{+}^{(3)}+e^{-\eta/2}\sigma_{+}^{(2)}\sigma_{-}^{(3)}\right)\text{ exp}\left[-\eta \frac{\sigma_{z}^{(2)}}{4}\right]\text{ exp}\left[\eta \frac{\sigma_{z}^{(3)}}{4}\right]\cdot  \notag \\
    &\cdot\text{ exp}\left[-\eta \frac{\sigma_{z}^{(1)}}{2}\right]+\left[\sum_{i=1}^{3}\frac12 \sigma_{z}^{(i)}\right]_{q}\left[\sum_{s=1}^{3} \frac12 \sigma_{z}^{(s)}+\mathbf{I}^{(3)}\right]_{q}-3\,[1/2]_q\,[3/2]_q\Biggl),
    \label{qexchange3}
\end{align}
which is not a straighforward superposition of $q$-exchange operators~\eqref{finalqKSpmpauliN2q}, and is indeed much more involved than its underformed counterpart
\be 
    H_{KS}^{(3)} =- I 
    \left(\Vec{\sigma}_{1}\cdot\Vec{\sigma}_{2} + \Vec{\sigma}_{1}\cdot\Vec{\sigma}_{3} + \Vec{\sigma}_{2}\cdot\Vec{\sigma}_{3} 
    \right).
\ee

The eigenstates can be obtained using the $q$-Clebsch--Gordan coefficients~\eqref{eq:qcg} to compose the $J=1$ and $J=0$ representations with an additional spin $1/2$ representation.
As a result, in the ferromagnetic case the ground states generate a $J=3/2$ quadruplet, given by the $q$-Dicke states
\begin{equation}
\begin{split}
\ket{3/2,-3/2}&= |\hspace{-0.1cm}\down\down\down\rangle,\qquad \ket{{3/2,-1/2}}=\frac{1}{\sqrt{[3]_q}}\left( q^{1/2}|\hspace{-0.1cm}\down\down\up\rangle +|\hspace{-0.1cm}\down\up\down\rangle+ q^{-1/2}|\hspace{-0.1cm}\up\down\down\rangle\right), \\
\ket{{3/2,1/2}}&=\frac{1}{\sqrt{[3]_q}}\left( q^{1/2} |\hspace{-0.1cm}\down\up\up\rangle +|\hspace{-0.1cm}\up\down\up\rangle + q^{-1/2}|\hspace{-0.1cm}\up\up\down\rangle \right),\qquad \ket{{3/2,3/2}}= |\hspace{-0.1cm}\up\up\up\rangle  ,
\end{split}
\label{eq:3q3/2}
\end{equation}
in which again $q$
 governs the entanglement properties of the $q$-KS eigenstates. The four (degenerate) excited states are provided by two $J=1/2$ doublets, given by
\begin{equation}
\begin{split}
\ket{1/2,-1/2}_1&=\frac{q^{-1/4}}{\sqrt{[2]_{q} [3]_q}}\left(q^{1/2} |\hspace{-0.1cm}\up\down\down\rangle +q  |\hspace{-0.1cm}\down\up\down\rangle  -[2]_{q} |\hspace{-0.1cm}\down\down\up\rangle\right),
	\\
\ket{{1/2,1/2}}_1&=\frac{q^{1/4}
   }{\sqrt{[2]_{q} [3]_q}}\left([2]_{q}  |\hspace{-0.1cm}\up\up\down\rangle -q^{-1}|\hspace{-0.1cm}\up\down\up\rangle   - q^{-1/2} |\hspace{-0.1cm}\down\up\up\rangle\right),
\end{split}
\label{eq:3q1/2}
\end{equation}
and 
\begin{equation}
\begin{split}
\ket{1/2,-1/2}_2&=\frac{1}{ \sqrt{[2]_{q}}}\left( q^{1/4} |\hspace{-0.1cm}\up\down\down\rangle -q^{-1/4}|\hspace{-0.1cm}\down\up\down\rangle \right),
	\\
\ket{1/2,1/2}_2&=\frac{1}{ \sqrt{[2]_{q}}}\left( -q^{-1/4} |\hspace{-0.1cm}\down\up\up\rangle +q^{1/4}|\hspace{-0.1cm}\up\down\up\rangle \right) .
\end{split}
\label{eq:3q1/20}
\end{equation}
These two doublets span the subspace of the ground states in the antiferromagnetic case. {Note that other basis for the eigenstates of the $q$-deformed model could be used, but this is the one directly given by the $q$-Clebsch--Gordan coefficients.}

{We emphasise that expressions~\eqref{finalqKSpmpauliN2q} and \eqref{qexchange3} show that the $q$-KS model is by no means equivalent to the spin-$1/2$ XXZ Heisenberg chain on the complete graph, since the latter would be given by a Hamiltonian of the type~\cite{bjornberg2020quantum}}
{
\begin{equation}
H=-\frac I N  \sum_{i<j}^{N}({S_x}^{(i)}{S_x}^{(j)}+ {S_y}^{(i)}{S_y}^{(j)} + \delta \,{S_z}^{(i)}{S_z}^{(j)})\, , 
\qquad
\delta\in [-1,1] \, ,\quad  \delta\neq 0\, .
    \label{eq:XXZcg}
\end{equation}
}
%{This is fully consistent with the fact that the $U_q(\mathfrak{su}(2))$ coalgebra symmetry used to construct the $q$-KS model strongly differs from the $U_q(\mathfrak{su}(2))$ symmetry underlying the anisotropy of he spin-$1/2$ Heisenberg XXZ Hamiltonian with interactions between nearest-neighbours.}

%%%%%%%%%

\section{Higher spin $q$-KS models}

The generic $q$-KS model~\eqref{qKScoalgebrah0} is expressed in terms of $U_q(\mathfrak{su}(2))$ generators, which only coincide with $U(\mathfrak{su}(2))$ spin operators in the $j=1/2$ representation. However, higher-spin $q$-KS models can be obtained using the known relationships between both sets of generators for arbitrary $j$.

\subsection{The $q$-KS model for $j=1$}

For $j=1$, the fundamental representation of the undeformed angular momenta operators is obtained by applying~\eqref{irepsu2}, namely 
\begin{equation}
J_+ = 
\begin{pmatrix}
0 & \sqrt{2} & 0 \\
0 & 0 & \sqrt{2} \\
0 & 0 & 0 \\
\end{pmatrix}
\equiv \Sigma_+, \qquad
J_- =
\begin{pmatrix}
0 & 0 & 0 \\
\sqrt{2} & 0 & 0 \\
0 & \sqrt{2} & 0 \\
\end{pmatrix}
\equiv \Sigma_-, \qquad
J_z = 
\begin{pmatrix}
1 & 0 & 0 \\
0 & 0 & 0 \\
0 & 0 & -1 \\
\end{pmatrix}
\equiv \Sigma_z.
\label{spin1}
\end{equation}
In the $q$-deformed case, the $j=1$ representation~\eqref{repsuq2} of the $U_q(\mathfrak{su}(2))$ generators is given by 
\begin{equation}
L_+ =
\begin{pmatrix}
0 & \sqrt{[2]_{q}} & 0 \\
0 & 0 & \sqrt{[2]_{q}} \\
0 & 0 & 0 \\
\end{pmatrix}, \qquad
L_- =
\begin{pmatrix}
0 & 0 & 0 \\
\sqrt{[2]_{q}} & 0 & 0 \\
0 & \sqrt{[2]_{q}} & 0 \\
\end{pmatrix}, \qquad
L_z = 
\begin{pmatrix}
1 & 0 & 0 \\
0 & 0 & 0 \\
0 & 0 & -1 \\
\end{pmatrix},
\end{equation}
and we realise that the quantum group generators are just proportional to the non-deformed $j=1$ spin operators~\eqref{spin1}:
\begin{equation}
 L_{\pm}=\sqrt{\frac{[2]_q}{2}}\Sigma_{\pm},
 \qquad
 L_z=\Sigma_z \, .
    \label{relation}
\end{equation}
Therefore, by substituting~\eqref{relation} in the generic expression of the $q$-KS model given in~\eqref{finalqKSpm}, we obtain the Hamiltonian  which defines a new exactly solvable model for $N$ particles with a spin $j=1$. 

Eigenvalues of this model are given by 
\be
E_q(j_{\alpha_i},m_{\alpha_i})=-\frac{I}{2}\left( [j_{\alpha_i}]_q \, [j_{\alpha_i}+1]_q-
N\,[2]_q
\right) -\gamma\,h\, m_{\alpha_i} ,
\label{eigenvaluesqKS1}
\ee
and the irreducible blocks for $N$ particles with spin $j=1$ will have a total spin given by $j_{\alpha_i}=N, N-1,\dots, 0$. The corresponding degeneracies $d_{NS}$ can be found in~\cite{VanVleck,curtright2017spin}, and the partition function is again given by the formula~\eqref{partition}.

\subsection{The $q$-KS model for arbitrary spin}

For any spin $j>1$, the $q$-KS model can also be written in terms of non-deformed physical spin-$j$ undeformed operators by using the so-called `deforming functional' approach to the $U_q(\mathfrak{su}(2))$ representation theory~\cite{curtright1990deforming,fairlie1990quantum,ballesteros1992characterization}. In particular, if we impose $L_\pm$ to be Hermitian conjugate, the $U_q(\mathfrak{su}(2))$ generators $L$ can be written in terms of the $U(\mathfrak{su}(2))$ physical spin operators $J$ as:
\be
 L_3 =J_3 \, ,\qquad
 L_+ =J_+ \ \sqrt{\frac{[j-J_3]_q [j+J_3+1]_q}{(j-J_3) (j+J_3+1)}} \, ,\qquad
L_- =\sqrt{\frac{[j-J_3]_q [j+J_3+1]_q}{(j-J_3) (j+J_3+1)}}\ J_-\, ,
\label{fr}
\ee
and the $q$-KS model for spin $j$ is straightforwardly obtained by inserting~\eqref{fr} into ~\eqref{finalqKSpm}.
The resulting expression is quite involved, but it provides a family of exactly solvable models for any spin $j$ whose spectra and eigenvectors would be deduced again using the representation theory of $U_q(\mathfrak{su}(2))$.

%%%%%%%%%%%%%%%%%%

\section{Curie temperature for the $q$-KS model}
{As a first step in the investigation of the thermodynamic properties of the model, we study in the following the Curie temperature for both undeformed and deformed KS models with spin-$1/2$.
In the $q$-KS case, we observe a completely different behaviour of the contribution of each energy level to the partition function, thus precluding the analytical approach proposed in~\cite{kittel2018introduction}. This forces us to  propose a new numerical method in order to obtain a $q$-dependent Curie temperature, which will be shown to be consistent with the observed dependence of the magnetization phase transition in terms of $T$.}

\subsection{Undeformed KS model}

{In the approach described in \cite{kittel1965development}, the Curie temperature is determined for a system of $j=1/2$ spins by finding the energy level that maximises its contribution to the partition function, which applies to large numbers of particles. Using Stirling's approximation and neglecting subleading terms, such a level is found as the solution of a transcendental equation. The Curie temperature thus arises as the critical point where a nontrivial solution of the equation exists, namely}
{\begin{equation}
    T_{C}=\frac{IN}{4\,k_{B}}\, ,
    \label{KittelSD}
\end{equation}
}
{under which the spontaneous magnetisation occurs. This is consistent with the mean field approach (see for instance~\cite{bjornberg2020quantum,kittel2018introduction}) that leads to the following Curie temperature}
{\begin{equation}
    T_{C}=\frac{INj(j+1)}{3k_{B}} \, ,
\end{equation}
}
for a KS model with an arbitrary spin $j$. Note that, in both cases, in the thermodynamic limit, we have to take $I\rightarrow I/N$, and then the Curie temperature becomes independent of $N$. We recall that in the realm of statistical physics, the occurrence of phase transitions is influenced by the dimensionality of the system ~\cite{le2004equilibrium}. {Since in the Kittel--Shore model the energy scales with $N^2$, the interaction constant $I$ must be replaced by $I/N$ in the thermodynamic limit, ensuring the extensibility of the model for any number of spins~\cite{le2004equilibrium}.}

{Unfortunately, in the $q$-deformed case, we have not been able to implement the analytical techniques giving rise to the previous results to obtain an analytic expression for the Curie temperature. Nevertheless, an alternative numerical approach can be developed for the spin-$1/2$ case by examining the  behaviour of the relative contribution of each energy level to the partition function in the undeformed model when $T$ approaches the phase transition and by generalising this analysis in the case of the $q$-KS model}. 

{In particular, let us write the partition function~\eqref{partition1} of the undeformed KS model in the form}
{\begin{equation}
    Z_N=\sum_{i}^{} z(E_i) = \sum_{p=0}^{p_m} z_p \, ,
\end{equation}
}
{where $z(E_i)$ is the contribution to the partition function of the states with energy $E_i$, which can also be labelled by the quantum number $p$, where $p_m=N/2$ if $N$ is even and $p_m=(N-1)/2$ if $N$ is odd (see Section 2.3).  This is shown in {Figure~\ref{fig6}}, where we plot each normalised contribution $z(E_i)/Z$ against its corresponding energy for $N=10^5$ and various temperatures. In general, we can check that the energy value $E_i$ for which the distribution of $z(E_i)/Z$ reaches its maximum value increases with $T$, and we have checked (through a large number of numerical computations) that the distinctive feature of the Curie temperature is precisely that for $T_C$ the maximum value of $z(E_i)/Z$ is smaller than the maximum of the distribution for any other $T$, as it can be appreciated in {Figure~\ref{fig6}}.
As we will see in the following, this kind of analysis of the shape of the partition function distribution will provide good numerical estimates of the Curie temperature for the $q$-KS model.}

\begin{figure}[H]
%\centering
  \begin{minipage}{0.4\textwidth}
    %\centering
    (a)\includegraphics[scale=0.7]{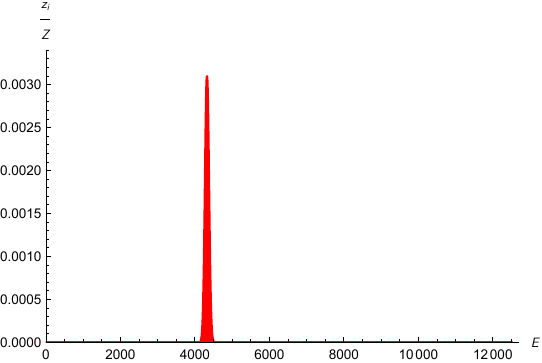}
  \end{minipage}
  \hspace{15mm}
  \begin{minipage}{0.4\textwidth}
    %\centering
    (b)\includegraphics[scale=0.7]{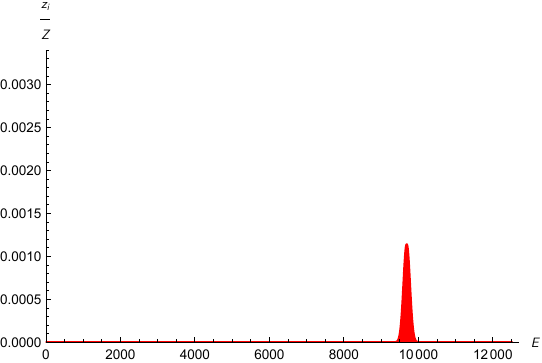}
  \end{minipage}
\end{figure}
\begin{figure}[H]
%\centering
  \begin{minipage}{0.4\textwidth}
    %\centering
    (c)\includegraphics[scale=0.7]{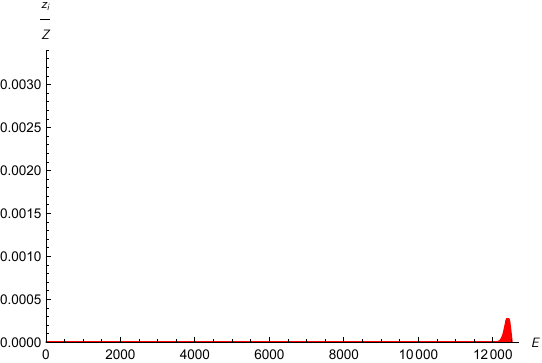}
  \end{minipage}
  \hspace{15mm}
  \begin{minipage}{0.4\textwidth}
    %\centering
    (d)\includegraphics[scale=0.7]{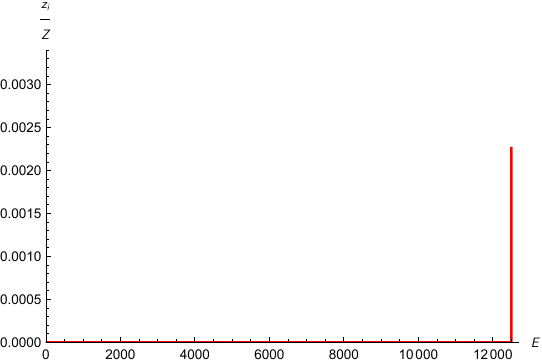}
  \end{minipage}
  \caption{Normalized weight of each energy level in the partition function for $N=10^5$  and different temperatures, in the ferromagnetic case with $I=1/N$. (a) $T=0.18$, (b) $T=0.23$, (c) $T=T_{C}=0.25$ and (d) $T=0.4$.}
  \label{fig6}
\end{figure}

\subsection{Deformed KS model}

{As we have seen in Section 3, in the $q$-KS model, the number of different energy levels and their degeneracies fully coincide with those of the undeformed model for any value of the deformation parameter $q=e^\eta$. However, the energy eigenvalues~\eqref{eigenvaluesqKS12} are now given by $q$-numbers, which grow exponentially for large values of $\eta$, as shown in Figure~\ref{qkskj1/2eigenD}. This will be the essential feature that  induces strong differences within the structure of the contributions to the $q$-deformed partition function in terms of the new energies.
Moreover, under $q$-deformation, the energy scales exponentially for $N$ large, and the simplest form to assure the extensibility of the model for an arbitrary $N$ consists of defining the coupling constant of the deformed model as $I\rightarrow I/N$ (as in the undeformed KS model) and the  deformation parameter as $\eta\rightarrow\eta/N$. With the latter change, the energy now scales as $N^2$, so {the extensibility of the model is guaranteed for any value of $\eta$ and any number of spins}. Therefore, in the following we will consider the $q$-KS model with deformation parameter $q=e^{\eta/N}$.}

{As a first step, we must verify that a phase transition for the magnetisation also occurs in the $q$-KS model. This is indeed the case, as shown in Figure~\ref{chilimitdeformed} where the magnetic susceptibility (\ref{eq:thermody}) for the $q$-deformed model is represented for different values of $q=e^{\eta/N}$, including the undeformed ($\eta=0$) case. In this way, these graphs (for a sufficiently large $N$) can be used to provide a numerical estimate of the Curie temperature $T_C(\eta)$ for different values of the deformation parameter $\eta$.} 

\begin{figure}[H]
%\centering
  \begin{minipage}{0.4\textwidth}
    %\centering
    (a)\includegraphics[scale=0.7]{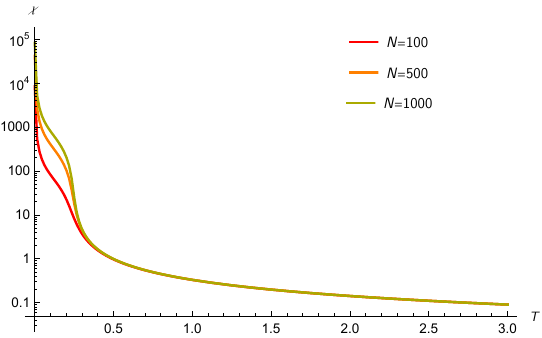}
  \end{minipage}
  \hspace{15mm}
  \begin{minipage}{0.4\textwidth}
    %\centering
    (b)\includegraphics[scale=0.7]{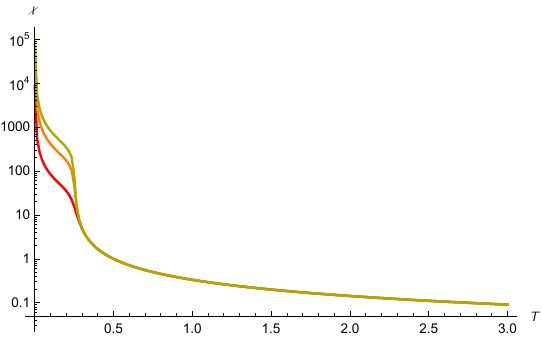}
  \end{minipage}
\end{figure}
\begin{figure}[H]
%\centering
  \begin{minipage}{0.4\textwidth}
    %\centering
    (c)\includegraphics[scale=0.7]{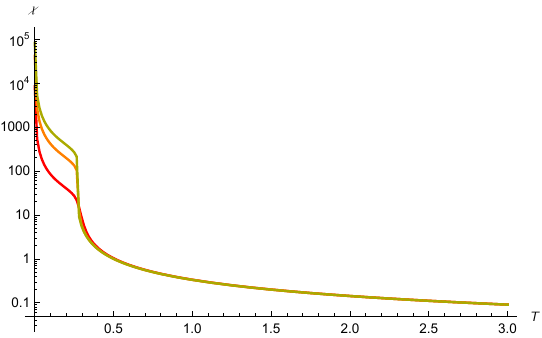}
  \end{minipage}
  \hspace{15mm}
  \begin{minipage}{0.4\textwidth}
    %\centering
    (d)\includegraphics[scale=0.7]{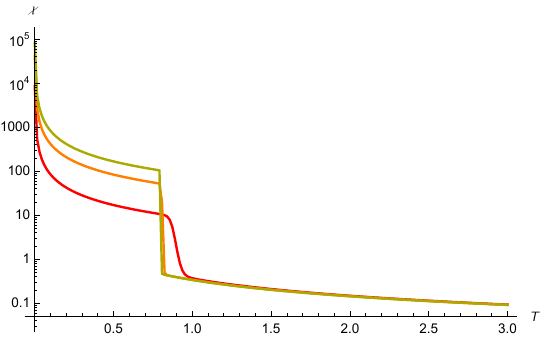}
  \end{minipage}
  \caption{Magnetic susceptibility of the $q$-KS model (in the ferromagnetic case) as a function of temperature for $N=100$ (red), $500$ (orange), and $1000$ (yellow). The values of the parameter $q=e^{\eta/N}$ are: (a) $\eta=0$, (b) $\eta=3$, (c) $\eta=4$, and (d) $\eta=9$. Note that the Curie temperature becomes sharply defined as long as $N$ increases, as expected.}
  \label{chilimitdeformed}
\end{figure}

{In Figure~\ref{chilimitdeformed}, we notice that as the deformation parameter $\eta$ rises, $T_C(\eta)$ also increases. In fact, the Curie temperature, obtained numerically as the transition temperature from the corresponding susceptibility plots, is represented in Figure~\ref{Cureta} for $N=10^6$ as a function of $\eta$. This logarithmic plot shows an essentially exponential dependence of $T_C$ in terms of $\eta$ for large values of the latter.} 
\begin{figure}[H]
    \centering
    \includegraphics[scale=1.0]{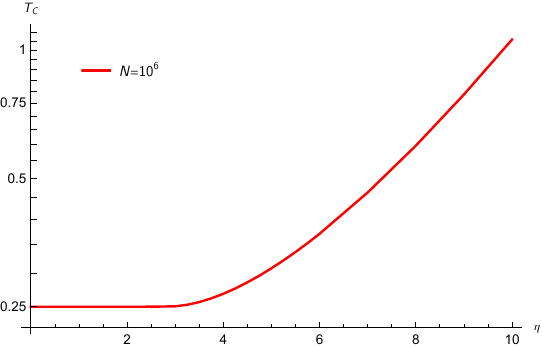}
    \caption{Curie temperature of the $q$-KS model as a function of the deformation parameter $\eta=\log q$ for $N=10^6$.}
    \label{Cureta}
\end{figure}

{In order to  obtain an analytical estimate of the function $T_C(\eta)$ represented in Figure~\ref{Cureta}, we consider the partition function of the $q$-KS model in the form}
{\begin{equation}
    Z_N^\eta=\sum_{i}^{} z^\eta(E^\eta_i) =\sum_{p=0}^{p_m} z_p^\eta \, ,
\end{equation}
}
{and perform an exhaustive numerical analysis of the distribution of the normalized contributions $z^\eta(E_i)/Z_N^\eta$ for different temperatures and deformation parameters.} 

{As a first illustrative example, Figure~\ref{fpev} shows the normalised contributions $z^\eta(E^\eta_i)/Z_N^\eta$ for $N=100$ and different values of $\eta$, considering always the Curie temperature $T_C(\eta)$ arising numerically from the discontinuity in the magnetic susceptibility. Larger values of $\eta$ widen the distribution $z^\eta(E^\eta_i)/Z_N^\eta$, first creating a plateau (Figure~\ref{fpev}-c) and, once the deformation parameter $\eta$ reaches larger values, two different peaks appear (Figure~\ref{fpev}-d). Numerical experiments for different $N$ and $\eta$ confirm this very same pattern and, therefore, we can conclude that the argument used in the non-deformed model in order to identify the Curie temperature in terms of the peak of the distribution $z^\eta(E^\eta_i)/Z_N^\eta$ cannot be applied for sufficiently large deformation parameters, since such a unique peak does not exist.}

\begin{figure}[H]
%\centering
  \begin{minipage}{0.4\textwidth}
    %\centering
    (a)\includegraphics[scale=0.7]{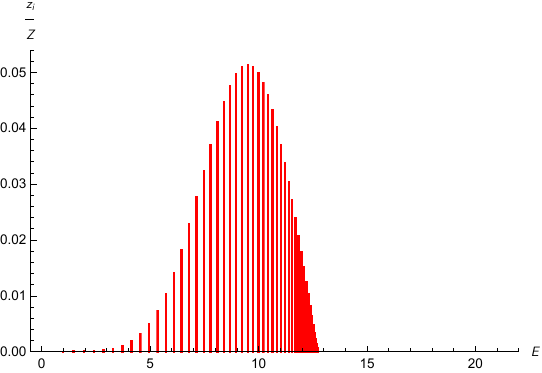}
  \end{minipage}
  \hspace{15mm}
  \begin{minipage}{0.4\textwidth}
    %\centering
    (b)\includegraphics[scale=0.7]{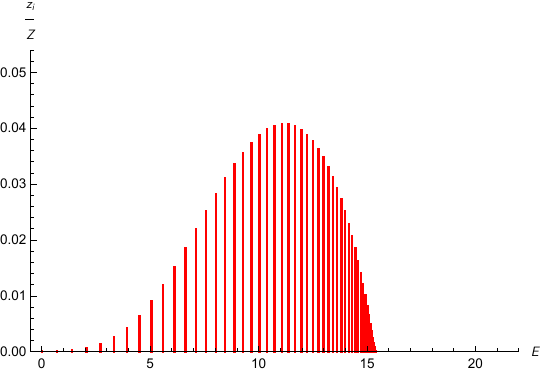}
  \end{minipage}
\end{figure}
\begin{figure}[H]
%\centering
  \begin{minipage}{0.4\textwidth}
    %\centering
    (c)\includegraphics[scale=0.7]{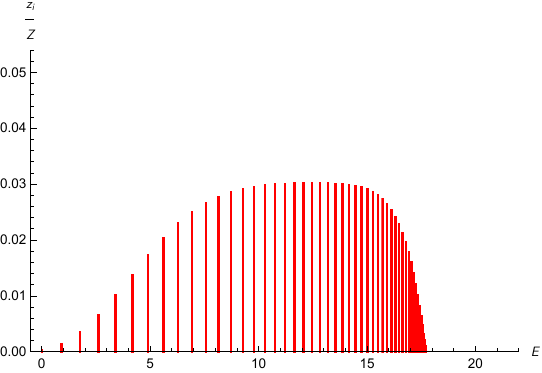}
  \end{minipage}
  \hspace{15mm}
  \begin{minipage}{0.4\textwidth}
    %\centering
    (d)\includegraphics[scale=0.7]{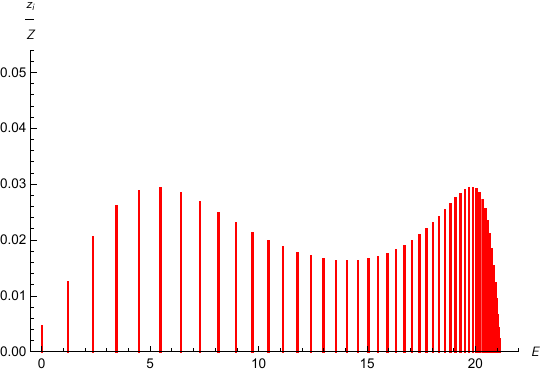}
  \end{minipage}
  \caption{Normalized contribution of each energy level $E_i$ in the partition function for number of particles $N=100$, deformation parameter $q=e^{\eta/N}$ for (a) $\eta=0$, (b) $\eta=3$, (c) $\eta=4$, (d) $\eta=5$ and Curie temperature $T_C(\eta)$ given by (a) $T=T_{C}(2)=0.250$, (b) $T=T_{C}(3)=0.270$, (c) $T=T_{C}(4)=0.294$, (d) $T=T_{C}(5)=0.340$ (ferromagnetic case).}
  \label{fpev}
\end{figure}

{This behaviour is confirmed in  Figure~\ref{fpdef2}, which shows the evolution for different temperatures of the distribution $z^\eta(E^\eta_i)/Z_N^\eta$ with $N=100$ and a fixed (large) deformation parameter $\eta=9$. Again, two maxima appear in the distribution when $T$ is sufficiently large, and as $T$ increases, the dominant maximum of the distribution appears at the highest energies. Remarkably, the Curie temperature determined from the susceptibility is represented in Figure~\ref{fpdef2}-c, where both peaks in the distribution $z^\eta(E^\eta_i)/Z_N^\eta$ (the one for $E_i=0$ and the one close to the maximum energy) have exactly the same value {to each other}. To ensure that this phenomenon persists in the thermodynamic limit, we display in Figure~\ref{fpdef3} the corresponding distributions for $N=10^5$, and we get the same pattern.}

\begin{figure}[H]
%\centering
  \begin{minipage}{0.4\textwidth}
    %\centering
    (a)\includegraphics[scale=0.7]{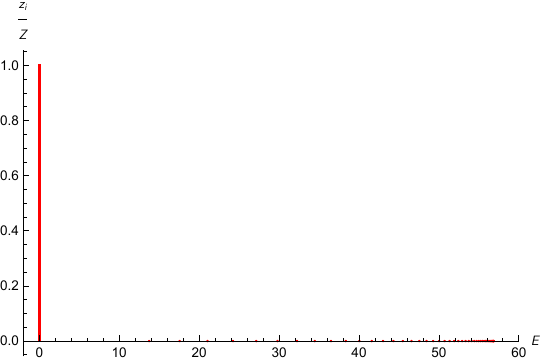}
  \end{minipage}
  \hspace{15mm}
  \begin{minipage}{0.4\textwidth}
    %\centering
    (b)\includegraphics[scale=0.7]{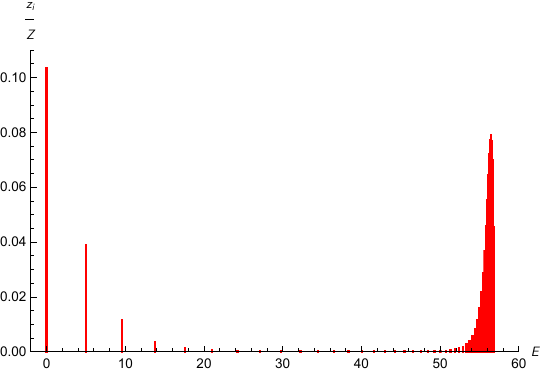}
  \end{minipage}
\end{figure}
\begin{figure}[H]
%\centering
  \begin{minipage}{0.4\textwidth}
    %\centering
    (c)\includegraphics[scale=0.7]{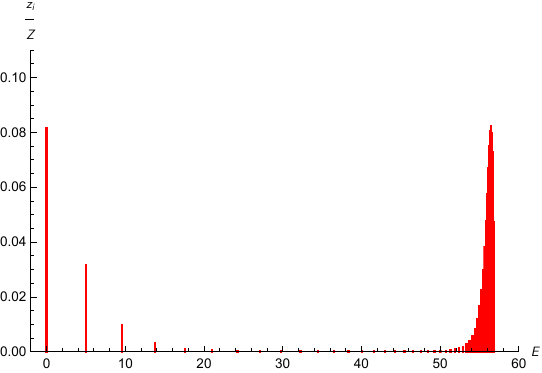}
  \end{minipage}
  \hspace{15mm}
  \begin{minipage}{0.4\textwidth}
    %\centering
    (d)\includegraphics[scale=0.7]{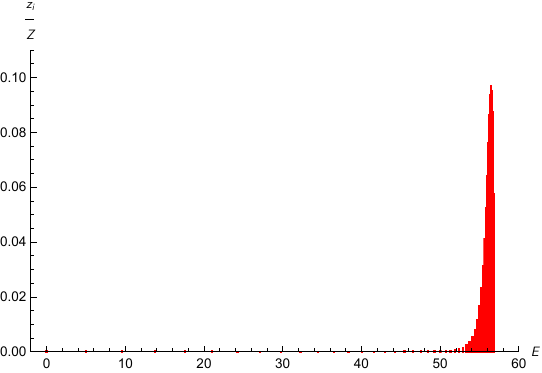}
  \end{minipage}
  \caption{Normalized weight of each energy level in the partition function for number of particles $N=100$, deformation parameter $\eta=9$ and temperatures (a) $T=0.250$, (b) $T=0.900$, (c) $T=T_{C}(9)=0.904$, and (d) $T=1.000$, for the ferromagnetic case.}
  \label{fpdef2}
\end{figure}

\begin{figure}[H]
%\centering
  \begin{minipage}{0.4\textwidth}
    %\centering
    (a)\includegraphics[scale=0.7]{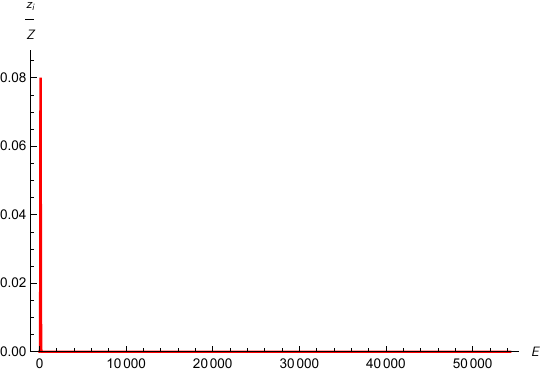}
  \end{minipage}
  \hspace{15mm}
  \begin{minipage}{0.4\textwidth}
    %\centering
    (b)\includegraphics[scale=0.7]{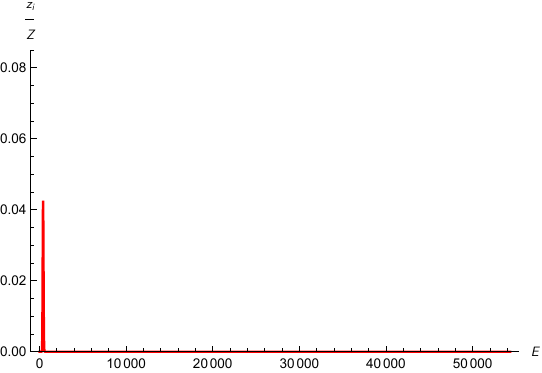}
  \end{minipage}
\end{figure}
\begin{figure}[H]
%\centering
  \begin{minipage}{0.4\textwidth}
    %\centering
    (c)\includegraphics[scale=0.7]{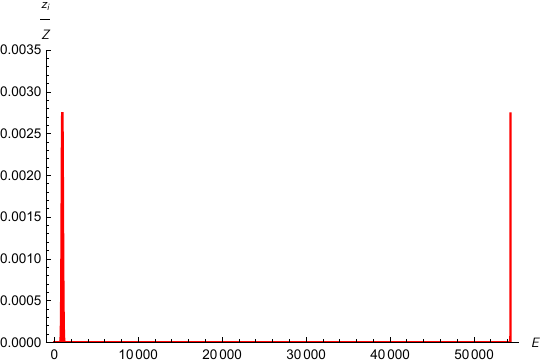}
  \end{minipage}
  \hspace{15mm}
  \begin{minipage}{0.4\textwidth}
    %\centering
    (d)\includegraphics[scale=0.7]{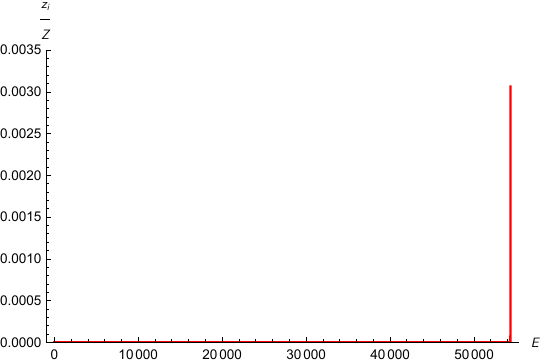}
  \end{minipage}
  \caption{Normalized weight of each energy level in the partition function for $N=10^5$, deformation parameter $\eta=9$ and temperatures (a) $T=0.6$, (b) $T=0.7$, (c) $T=T_{C}(9)=0.786189$, and (d) $T=0.8$, for the ferromagnetic case.}
  \label{fpdef3}
\end{figure}

{This behaviour can be explained by writing explicitly the partition function}
{\begin{equation}
    Z_N^{\eta}=\sum_{p=0}^{N/2}(N-2p+1)^{2}\frac{N!}{(N-p+1)!p!}\text{exp}\left[\frac{\beta I}{2N}\left(\left[\frac{N}{2}-p\right]_{q}\left[\frac{N}{2}-p+1\right]_{q}-N\left[\frac{1}{2}\right]_{q}\left[\frac{3}{2}\right]_{q}\right) +\beta\,E_0^{\eta}\right],
    \label{eq:qpart}
\end{equation} 
}
{where $q=e^{\eta/N}$ and $E_0^{\eta}=-\frac{ I}{2N}\left(\left[\frac{N}{2}\right]_{q}\left[\frac{N}{2}+1\right]_{q}-N\left[\frac{1}{2}\right]_{q}\left[\frac{3}{2}\right]_{q}\right)$. Here for a given $p$ (or the corresponding $E^\eta_i$), Figs.~\ref{fpev},~\ref{fpdef2} and \ref{fpdef3} reflect the competition within~\eqref{eq:qpart} between the factor describing the degeneracy of the levels with the same $E_i^\eta$ and the Boltzmann factor, where $T$ plays also a relevant role, and in which the exponent takes values in terms of $q$-numbers, which become very large as $\eta$ increases.}

{In order to derive an analytical estimation for the Curie temperature in terms of $\eta$, we have to take into account the qualitatively different behaviours of $Z_N^\eta$ for small and large deformation parameters. As Figure~\ref{fpev} illustrates, the partition function distinguishes two regimes: ``unimodal" (for small $\eta$) or ``bimodal" (large $\eta$). As Figure~\ref{Cureta} shows, in the thermodynamic limit the unimodal range corresponds approximately to $\eta\in(0,3)$ and the bimodal range to $\eta>3$.} 

{For small and intermediate deformation parameters ($\eta \in (0,8)$), a piecewise fit to the data in Figure~\ref{Cureta} yields the following functions:} 
{\begin{align}
    T_{C}(\eta)=&0.25 \hspace{12.25cm} \eta\in(0,3),   \label{tc1}\\
    T_{C}(\eta)=&0.25+0.022959\eta-0.023077\eta^2+0.007164\eta^3-0.000779\eta^4+0.000035\eta^5 \hspace{1cm} \eta\in(3,8),
    \label{tc2}
\end{align}
}
{where the correlation coefficient is $R^2=0.999998$ (and the root mean squared error is $0.00018$). Indeed, more precise fittings can be implemented.} 

{For larger values of $\eta$, we make use of a different approach based on the
exact expression for the q-deformed partition function~\eqref{eq:qpart}.  As seen in Figs.~\ref{fpdef2} and~\ref{fpdef3}, the Curie temperature occurs when the two maxima in the distribution $z^\eta(E^\eta_i)/Z_N^\eta$ coincide. Therefore, we can obtain the Curie temperature by imposing that the two maxima of $z^\eta(E^\eta_i)/Z_N^\eta$ take the same value. Note that for obtaining the Curie temperature analytically we just need to identify the energy levels at which both identical maxima occur.}

{It can be checked that the larger the deformation $\eta$ and the number of particles are, the {first} maximum occurs closer to $p=0$ ($E^\eta_i=0$) and the second one approaches to $p=N/2$ (the largest $E^\eta_i$). Then, we can assume that in the limiting case with large $\eta$ and $N$ the first maximum appears exactly for $p=0$ and the {second one} for $p=N/2$. By imposing in~\eqref{eq:qpart} that these two maxima take the same value, one finds the following expression:}
{\begin{equation}
    T_{C}(\eta)=\frac{I \left[\frac{N}{2}+1\right]_{q} \left[\frac{N}{2}\right]_{q}}{2 k_B N \log \left(\frac{N!}{(N+1)\left(\frac{N}{2}\right)!\left(\frac{N}{2}+1\right)!}\right)},
    \label{eq:TempCurieDef}
\end{equation}
}
{where, as usual, $q=e^{\eta/N}$. This provides an excellent approximation for Curie temperatures in systems with many particles and a large deformation parameter. Morever, by taking the thermodynamic limit of (\ref{eq:TempCurieDef}), we obtain}
{
\begin{equation}
    \lim_{N\rightarrow \infty} T_{C}(\eta)=\frac{2 \sinh ^2\left(\frac{\eta}{4}\right)}{\eta^2 \log (2)}\thicksim\frac{e^{\eta/2}}{\eta^2 \log (4)},
    \label{tc3}
\end{equation}
}
{which fits perfectly with Figure~\ref{Cureta} for $\eta>8$. Summarizing, the Curie temperature $T_{C}(\eta)$ for the $q$-KS model can be obtained using Eqs.~\eqref{tc1}, \eqref{tc2} and \eqref{tc3}.} 

%%%%%%%%%%%%%%%%%%%%%%%%%
\section{Concluding remarks}

The new integrable spin Hamiltonians presented here were constructed as integrable deformations of the KS models by imposing {\em ab initio} the preservation of a $q$-deformation of the underlying coalgebra symmetry of the latter. The second essential component that leads to the exact solvability of the $q$-KS models is the full structural coincidence between the theory of irreducible representations of $U_q(\mathfrak{su}(2))$ and that of $U(\mathfrak{su}(2))$ (when $q$ is not a root of unity). {It is worth stressing that integrable long-range interacting spin models like $q$-KS one here introduced are rare, and provide valuable tools in order to understand long-range phenomena, and thus allowing applications in different contexts (see~\cite{lamers2022spin,hakobyan1996spin,matushko2022elliptic,klabbers2024deformed} and references therein). Therefore, $q$-deformed analogues of these models are relevant since they preserve complete integrability and provide an additional parameter to play with in order to explore novel features and applications.}

From a mathematical perspective, an interesting open problem would be the construction of $q$-KS models for $q$ a root of unity. {Also, the case when $q$ is not a root of unity but belongs to the unit circle seems worth exploring, since in this case the eigenvalues and eigenfunctions presented here remain valid, and the relation of the associated nearest-neighbours XXZ Heisenberg model with conformal field theory through its $U_q(\mathfrak{su}(2))$ invariance is well-known~\cite{alcaraz1987surface,pasquier1990common,korff2007pt}}. Moreover, KS models (both deformed and undeformed) in which each particle $i$ has arbitrarily different spins $j_i$ could be considered and, in particular, spin models with two alternating spin values would be interesting. In all these cases, explicit formulas for the eigenvalues of the model rely on the knowledge of the explicit expressions for the multiplicities and eigenvalues of the corresponding tensor product representations, which is by no means an immediate task that, to the best of our knowledge, has not been considered so far in the literature. In addition, the study of the dynamics of the $q$-KS model for classical spins would be interesting from an integrability point of view and could be performed by following the analysis given in~\cite{magyari1987integrable} for the undeformed KS Hamiltonian and by applying the techniques developed in~\cite{ballesteros2003classical} for dynamical systems endowed with coalgebra symmetry. {Finally, it would be interesting to revisite the coalgebra symmetry approach for $U_q(\mathfrak{su}(2))$ in the framework of quantum Schur-Weyl duality and Hecke algebras~\cite{jimbo1986q} by exploring the connection between the set of commuting
Casimir operators $\Delta^{(m)}(C)$ with the quantum analogues of Jucys-Murphy elements, which generate the center of the group algebra of the symmetric group. In this direction, we recall that $q$-Dicke states have been used in~\cite{qpermutations} to show that the $q$-symmetric subspace can be described in terms of a non-unitary representation of the symmetric group.}

Obviously, a second pending task consists of the study of the physical properties {and the thermodynamical limit of the $q$-KS models, starting from the $j=1/2$ and $j=1$ cases.} All these properties are expected to behave as smooth deformations of those for the KS model {although, as the study of the Curie temperature $T_{C}(\eta)$ has shown, a careful analysis is needed and will be presented in~\cite{Future}.} In addition, the quantum informational perspective for both the KS and $q$-KS models is worth developing: entanglement entropies should be analysed and, in particular, the role of $q$-Dicke states in this context could be relevant. Finally, the possible applications of $q$-KS Hamiltonians for small $N$ as effective models for systems with few-body spin couplings should be addressed (for instance, in small magnetic clusters~\cite{ciftja1999equation} and coupled quantum dots~\cite{loss1998quantum,woodworth2006few}). All of these problems are currently under investigation and will be presented elsewhere.

\section*{Acknowledgements}

The authors acknowledge partial support from the grant PID2023-148373NB-I00 funded by MCIN/AEI/ 10.13039/501100011033/FEDER -- UE, and the Q-CAYLE Project funded by the Regional Government of Castilla y León (Junta de Castilla y León) and the Ministry of Science and Innovation MICIN through NextGenerationEU (PRTR C17.I1). I. Gutierrez-Sagredo thanks Universidad de La Laguna, where part of the work has been done, for hospitality and support. {V. Mariscal acknowledges support from Universidad de Burgos through a PhD grant. The authors are indebted to the Referees for their constructive criticisms and comments, which have contributed to improve the content of the paper.}

\end{document}